\documentclass[prd,twocolumn,showpacs,reprint,preprintnumbers,nofootinbib]{revtex4-2}

\RequirePackage[colorlinks=true
,urlcolor=blue
,anchorcolor=blue
,citecolor=blue
,filecolor=blue
,linkcolor=blue
,menucolor=blue
,linktocpage=true
,pdfproducer=medialab
,pdfa=true
]{hyperref}

\usepackage{amsmath,amssymb,amsthm,amsfonts}
\usepackage{graphicx,tabularx}
\usepackage{color}
\usepackage{multirow}  
\usepackage{comment}
\usepackage{slashed}
\usepackage{enumitem}
\usepackage{cleveref}
\usepackage{rotating}
\usepackage[english]{babel}
\usepackage[justification=centerlast]{caption}
\usepackage{subcaption}
\usepackage{float}

\usepackage{cases}

\crefname{section}{Sec.}{Secs.}
\crefname{figure}{Fig.}{Figs.}
\crefname{equation}{Eq.}{Eqs.}
\crefname{appendix}{Appendix}{Appendices}
\setlist[description]{leftmargin=0.4cm}
\setlist[itemize]{leftmargin=0.4cm}

\newcommand{\be}{\begin{equation}\begin{aligned}}
\newcommand{\ee}{\end{aligned}\end{equation}}
\newcommand{\beq}{\begin{equation}}
\newcommand{\eeq}{\end{equation}}
\newcommand{\beqa}{\begin{eqnarray}}
\newcommand{\eeqa}{\end{eqnarray}}

\newcommand{\newc}{\newcommand*}

\long\def\begincomment#1\endcomment{%
        \begingroup\sf\baselineskip12pt#1\endgroup}

\newc{\etal}{\textrm{et al.}}
\newc{\eg}{\textrm{e.g.}}
\newc{\ie}{\textrm{i.e.}}
\newc{\etc}{\textrm{etc}}
\newc\vs{\textrm{vs.}}
\newc{\cl}{\rm {C.L.}}

\newc{\ev}{\ensuremath{\,\mathrm{eV}}}
\newc{\kev}{\ensuremath{\,\mathrm{keV}}}
\newc{\mev}{\ensuremath{\,\mathrm{MeV}}}
\newc{\gev}{\ensuremath{\,\mathrm{GeV}}}
\newc{\tev}{\ensuremath{\,\mathrm{TeV}}}
\newc{\MeV}{\mev}
\newc{\TeV}{\tev}
\newc{\invpb}{\ensuremath{/\text{pb}}}
\newc{\invfb}{\ensuremath{\,\text{fb}^{-1}}}
\newc\nb{\ensuremath{\,\mathrm{nb}}} \newc\pb{\ensuremath{\,\mathrm{pb}}} \newc\fb{\ensuremath{\,\mathrm{fb}}}
\newc\pc{\ensuremath{\,\mathrm{pc}}}
\newc\kpc{\ensuremath{\,\mathrm{kpc}}}
\newc\mpc{\ensuremath{\,\mathrm{Mpc}}}
\newc\ps{\ensuremath{\,\mathrm{ps}}}

\newc\cmeter{\ensuremath{\,\mathrm{cm}}}
\newc\meter{\ensuremath{\,\mathrm{m}}}
\newc\kmeter{\ensuremath{\,\mathrm{km}}}
\newc\second{\ensuremath{\,\mathrm{s}}}
\newc\msecond{\ensuremath{\,\mathrm{ms}}}
\newc\nsecond{\ensuremath{\,\mathrm{ns}}}
\newc\psecond{\ensuremath{\,\mathrm{ps}}}

\renewcommand{\eqref}[1]{Eq.~(\ref{#1})}

\RequirePackage[normalem]{ulem}

\makeatletter
\def\l@subsubsection#1#2{}
\makeatother

\DeclareUnicodeCharacter{2212}{\textendash}

\makeatletter
\let\cref@old@eq@setnumber\eq@setnumber
\def\eq@setnumber{%
\cref@old@eq@setnumber%
\cref@constructprefix{equation}{\cref@result}%
\protected@xdef\cref@currentlabel{%
[equation][\arabic{equation}][\cref@result]\p@equation\theequation}}
\makeatother

\usepackage{multirow}
\usepackage{makecell}

\usepackage{fontawesome}

\begin{document}

%============================================================
\title{Indirect detection of long-lived particles in a rich dark sector with a dark vector portal}
%============================================================

\author{Krzysztof~Jod\l{}owski}
\email{k.jodlowski@ibs.re.kr}
\affiliation{Particle Theory and Cosmology Group\char`,{} Center for Theoretical Physics of the Universe\char`,{} Institute for Basic Science (IBS)\char`,{} Daejeon\char`,{} 34126\char`,{} Korea}
\affiliation{National Centre for Nuclear Research, Pasteura 7, 02-093 Warsaw, Poland}

\author{Leszek~Roszkowski}
\email{leszek.roszkowski@ncbj.gov.pl}
\affiliation{Astrocent, Nicolaus Copernicus Astronomical Center Polish Academy of Sciences,ul. Bartycka 18, 00-716 Warsaw, Poland}
\affiliation{National Centre for Nuclear Research, Pasteura 7, 02-093 Warsaw, Poland}

\author{Sebastian~Trojanowski}
\email{s.trojanowski@ncbj.gov.pl}
\affiliation{Astrocent, Nicolaus Copernicus Astronomical Center Polish Academy of Sciences,ul. Bartycka 18, 00-716 Warsaw, Poland}
\affiliation{National Centre for Nuclear Research, Pasteura 7, 02-093 Warsaw, Poland}

\begin{abstract}
    Simplified models of light new physics provide a convenient benchmark for experimental searches for new physics signatures, including dark matter (DM). However, additional detection modes can arise in less simplified and more realistic scenarios where new degrees of freedom are invoked.
    In this study, we introduce a non-minimal model based on a popular dark photon portal to DM where the mediator mass is obtained by interactions with the dark Higgs boson which acts as a long-lived particle. 
    We further add to this scenario a new heavy DM species secluded from the Standard Model. In this model, which involves light and heavy particles in the dark sector, we find some new interesting phenomenological features that lead to complementary probes in intensity frontier searches for light long-lived particles, indirect detection searches for dark matter, and cosmic microwave background surveys. We also find possible non-local effects in the DM indirect detection searches that could significantly affect the usual detection strategies.
\end{abstract}

\renewcommand{\baselinestretch}{0.85}\normalsize
\maketitle
% \tableofcontents
\renewcommand{\baselinestretch}{1.0}\normalsize

%============================================================
\section{\label{sec:intro}Introduction}
%============================================================
The nature of the particle content of the dark matter (DM) sector of the Universe remains unknown. The dominant paradigm of the last several decades has been based on the assumption that DM is made up of some weakly interacting massive particles (WIMPs) in the approximate mass range of a few \gev\ to a few \tev. WIMP DM could naturally be produced in the early Universe via the popular thermal freeze-out mechanism, yielding a correct relic density for WIMP interactions set by a fraction of electroweak interactions of the Standard Model (SM); see, \eg, Ref.~\cite{Arcadi:2017kky,Roszkowski:2017nbc,Billard:2021uyg} for a review.\footnote{The other very popular and strongly motivated candidate is an axion; see, \eg, Ref.~\cite{Marsh:2015xka} for a review.} However, the lack of experimental signal in a wide array of worldwide searches has led one to explore alternative candidates for DM, its production mechanisms, and detection methods; see \eg, Refs~\cite{Battaglieri:2017aum,AlvesBatista:2021gzc}.

For example, the standard freeze-out mechanism can be generalized to broad ranges of DM mass and coupling constants~\cite{Feng:2008ya}; see also a discussion in Ref.~\cite{Baer:2014eja}. While too large values of DM couplings are constrained by unitarity, lowering them by even several orders of magnitude below the electroweak strength can still render the correct relic density, however, for DM mass much below the $\gev$ scale. Some light mediator particle(s) beyond the Standard Model (BSM) spectrum are present in various alternative scenarios. They specify the interaction between the SM and some DM particle with the mass in the $\gev$ range, or below~\cite{Boehm:2003hm,Pospelov:2007mp}. Experimental searches for such light DM (LDM) via direct detection (DD) invoke new  strategies~\cite{Battaglieri:2017aum,Billard:2021uyg}, while further signatures of associated frameworks may be possible with searches for unstable light mediator species; see Refs~\cite{Beacham:2019nyx,Alimena:2019zri,Agrawal:2021dbo,Antel:2023hkf} for recent reviews.

Such studies are usually performed within a simplified model framework where the DM particle is assumed to be part of some secluded (dark) sector that interacts with the SM sector only via one mediator particle, for instance, a dark photon or a dark Higgs boson; see Refs~\cite{Fabbrichesi:2020wbt,Filippi:2020kii,Arcadi:2021mag} for recent reviews. In more UV-complete models, further experimental probes could be possible and provide other complementary ways of their experimental tests, which triggers interest in considering non-minimal scenarios.

In models with a richer dark sector, there are typically more BSM particles with masses spanning a wide range of values, even up to several orders of magnitude. Suppose such a rich dark sector remains secluded from the SM sector, with only a weak portal communicating between them. In that case, the dark couplings of the BSM species are only mildly constrained and can lead to new phenomenological effects. This is especially relevant for cosmological tests and indirect detection (ID) searches of DM that can probe scatterings in the dark sector with the visible signals generated via subsequent decays of BSM species. Notable examples of such effects include, e.g., secluded~\cite{Pospelov:2007mp,Batell:2009yf} or Sommerfeld-enhanced~\cite{Bringmann:2016din} DM annihilations into the dark sector particles that modify ID bounds and can be constrained by Cosmic Microwave Background (CMB) radiation measurements.\footnote{Collider probes of light new physics can also be modified in the presence of large dark coupling constants via, e.g., possible secondary production of new BSM species in front of or inside the detectors~\cite{Jodlowski:2019ycu}.}

In this study, we focus on such an interplay between light and heavy dark species in models that can simultaneously be probed by experiments targeting signatures related to heavy DM and signals associated with light and long-lived particles (LLPs). As we will show, already in a simple extension of the most minimal scenarios in which we add a heavy DM candidate and much lighter secluded dark species, one can find new complementary detection prospects in future CMB surveys and  DM ID searches, in addition to standard LLP searches. 

We will focus on a popular dark photon portal, which is often considered to be a mediator between the SM and even very complicated BSM sectors~\cite{Pospelov:2007mp,Batell:2009di,Batell:2009yf,Fabbrichesi:2020wbt}. In particular, such a portal can naturally be related to the \textsl{hidden valley} scenarios predicting the existence of some LLPs~\cite{Strassler:2006im}. Motivated by this, we assume that the relevant new light species has a mass below $1\gev$ up to a few tens of GeV and suppressed couplings to the SM sector. The remaining non-minimal BSM content of our model will further contain particles at mass scales up to $10\tev$, or so, including a heavy secluded DM candidate that could be targeted in future ID observations.

As we will see, in our two-component DM scenario there are new effects that are not present in the minimal setup and that consequently lead to new experimental signatures. In particular, dark-bremsstrahlung annihilations of the dominant heavy DM particle might produce non-trivial spectra of SM species and be additionally affected by striking \textit{non-local DM ID} signatures that both require modified search strategies. The presence of very long-lived light mediator species might also leave imprints on CMB observations. In addition, in our model, accelerator-based intensity-frontier experiments could unveil the existence of light dark sector portal species. The combination of the above signatures clearly differs from simplified models with light, GeV-scale DM.

This paper is organized as follows. In \cref{sec:model}, we introduce the BSM model of our interest. In \cref{sec:relic}, we examine the relic abundance of both stable and long-lived dark species present in this scenario. In \cref{sec:constraintsandfuture}, we discuss past and future possible bounds on this model. Finally, we present our results in \cref{sec:results} and our conclusions in \cref{sec:conclusions}. The technical details of our analysis, including formulae for the relevant cross sections and decay widths, and a discussion of the photon spectrum in the DM ID, among others, have been moved to \cref{sec:sec_formulae,app:spectrum}.
In \cref{app:nonlocal}, we highlight possible unique non-local effects in DM ID that could appear in the presence of sufficiently long-lived light mediators.

%%%%%%%%%%%%%%%%%%%%%%%%%%%%%%%%%%%%%%%%%%%%%%%%%%%%
\section{Model\label{sec:model}}
%%%%%%%%%%%%%%%%%%%%%%%%%%%%%%%%%%%%%%%%%%%%%%%%%%%%

The BSM model that we put forward in this article contains a fairly rich dark sector that is coupled to the visible (SM) sector via the kinetic mixing between a dark photon $A^\prime$ and the SM hypercharge gauge boson. The dark photon is assumed to be a gauge boson of the secluded $U(1)^\prime$ group. Its mass is driven by the vacuum expectation value (vev) of a new scalar field, the dark Higgs boson $\sigma$, which is charged under $U(1)^\prime$. Additional light DM particle $\eta$ is often introduced in the model. It can be coupled to the SM via the dark photon or scalar portal. This is one of the prototype models predicting the existence of LLPs at the $\mev$-$\gev$ scale, which is discussed in intensity frontier studies~\cite{Pospelov:2007mp,Batell:2009di,Batell:2009yf}; see, \eg, Refs~\cite{Darme:2017glc,Darme:2018jmx,Duerr:2019dmv} for recent analyses.

\begin{figure}[tb]
\centering
\includegraphics[scale=0.28]{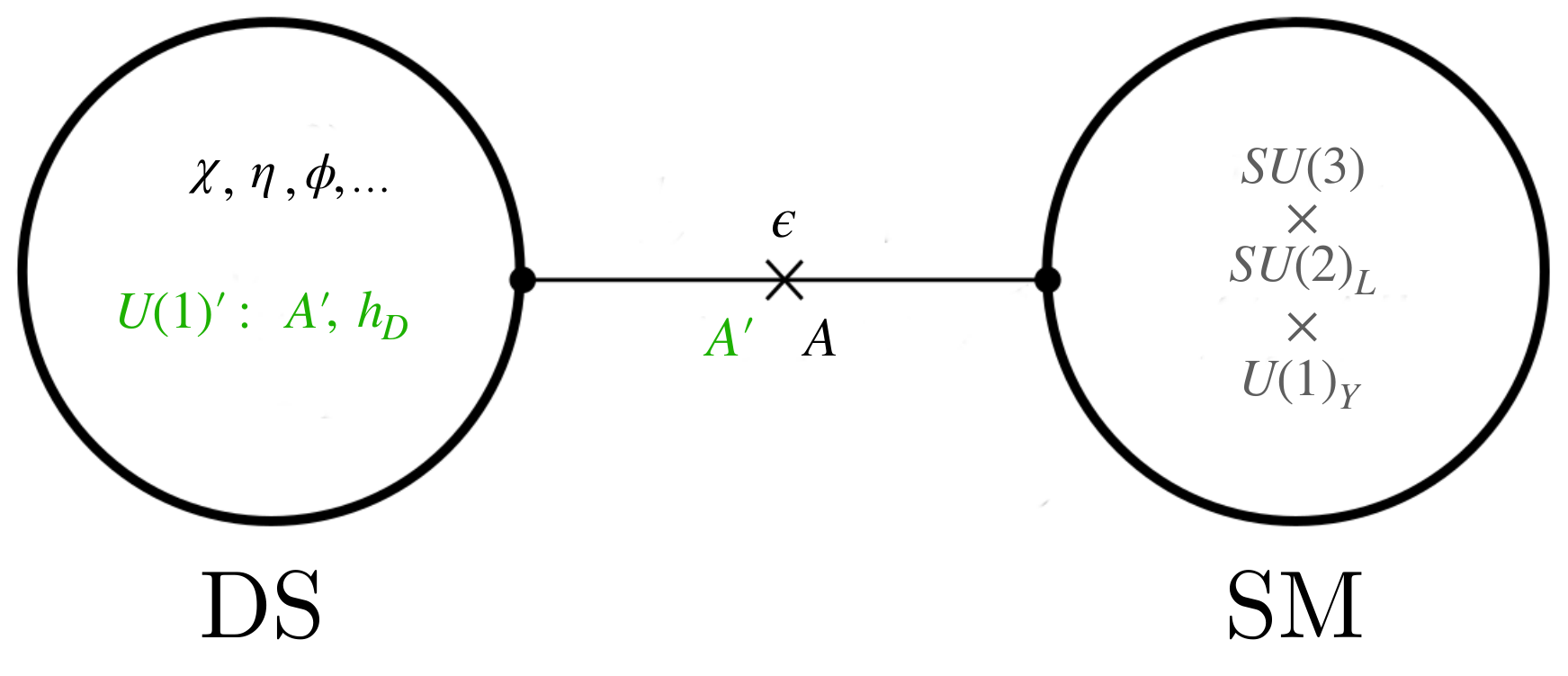}
\caption{
Schematic illustration of the model. The dark sector (DS) is connected to the Standard Model (SM) through a light dark photon portal ($A^\prime$), which kinetically mixes with the SM hypercharge gauge boson. We denote the mixing parameter by $\epsilon$. The mass of $A^\prime$ is determined by the vev of the additional dark Higgs boson $h_D$. The vector field and the dark Higgs field charged under the secluded $U(1)^\prime$ group are shown in green. The dark matter sector indicated with the black color consists of the additional scalar fields $\eta$ and $\chi$. The former mixes with $h_D$, while the latter is heavy and is the dominant DM component. It couples to $\eta$ via the auxiliary scalar field ($\phi$).
\label{fig:cartoon}}
\end{figure}

In our case, to illustrate the interplay between heavy and light new physics, we extend this popular scenario by adding one more heavy DM particle $\chi$. The dark matter sector of the model then comprises of two scalar fields $\eta$ and $\chi$. The former mixes with the dark Higgs boson. The latter has negligible interaction strength with all the light dark species besides $\eta$, to which it couples via an additional heavy scalar field $\phi$. This field only plays an auxiliary role in our analysis and could be replaced with a direct renormalizable contact operator, $|\chi|^2|\eta|^2$. The dominant heavy $\chi$ DM component is highly secluded from the SM, although it's still thermally produced in the early Universe. The light dark sector species guarantee thermal contact with the SM in a cascade process. A scenario of this type is phenomenologically distinct from traditional WIMPs, as we discuss below. In \cref{fig:cartoon}, we schematically present the connection between the SM and the dark sectors in the model. The fields charged under the dark gauge group $U(1)^\prime$ are shown in green.

The Lagrangian of the model can be written as
\begin{equation}
\mathcal{L}=\mathcal{L}_{\mathrm{SM}}+\mathcal{L}_{\mathrm{DS}}+\mathcal{L}_{\text {portal }}
\end{equation}
where $\mathcal{L}_{\mathrm{SM}}$ is the SM Lagrangian, $\mathcal{L}_{\mathrm{DS}}$ corresponds to the dark sector, and $\mathcal{L}_{\text {portal }}$ describes interactions between the SM and the dark sector
\begin{equation}
    \mathcal{L}_{\text {portal }} = -\frac{\epsilon}{2} F_{\mu \nu}^\prime F^{\mu \nu}.
\label{eq:L_portal}
\end{equation}
In the following, we assume $\epsilon\ll 1$ and typically $m_{A^\prime}<m_Z$ such that the mixing between the dark photon and the SM $Z$ boson can be neglected~\cite{Berlin:2018jbm}. We also assume, for simplicity, that the mixing between the dark and SM Higgs bosons is negligible; see Ref.~\cite{Lebedev:2021xey} for a recent review of the BSM Higgs portal. We parameterize both fields in the unitary gauge after the spontaneous symmetry breaking of both the electroweak and the dark gauge symmetries as follows
\begin{equation}
\Phi=\left(0,\left(v_{h}+h\right) / \sqrt{2}\right)^{T}, \quad \sigma=\left(v_{\mathrm{D}}+h_{\mathrm{D}}\right) / \sqrt{2}
\end{equation}
where the SM Higgs vacuum expectation value is $v_{h}=246\gev$, while the corresponding quantity for the dark Higgs field is denoted by $v_{D}$. We denote resulting physical eigenvector states by $H$ and $h_D$, respectively.

The following Lagrangian describes the interactions between the dark sector species 
\begin{align}
\label{eq:LagrDS}
\mathcal{L}_{\mathrm{DS}} \supset &\ \mu_{\chi} |\chi|^2 \phi + \mu_{\eta} |\eta|^2 \phi + (q^\prime_{h_D} g_D)^2 A^{\prime \mu} A^\prime_\mu |h_D|^2 .
\end{align}
As we will discuss below, invoking two components of DM scalar fields allows one to relax otherwise stringent bounds on heavy WIMP-like scalars coupled to the SM sector via light mediator species; see Refs~\cite{Ma:2017ucp,Duerr:2018mbd}. In the following, we set $q_{h_D}^\prime =1$, while we take $q^\prime_\eta = q^\prime_\chi = 0$. Hence, the $\eta$ DM component does not couple to the dark photon directly but only indirectly via the dark Higgs boson. The couplings of $\eta$ to the latter are described by the dark scalar potential terms 
\begin{align}
\mathcal{L}_{\mathrm{DS}}  \supset &\ \ \mu_{\mathrm{D}}^{2} |\sigma|^2-\frac{1}{2} \lambda_{\mathrm{D}}|\sigma|^4 \nonumber\\
&+m_{\chi}^{2}|\chi|^{2}+m_{\eta}^{2}|\eta|^{2}\nonumber\\
&-\lambda_{h_D \eta}h_D^2|\eta|^{2} -\left(\mu_{h_D \eta} h_D |\eta|^{2}+\mathrm{h.c.}\right).
\end{align}
As a result of the spontaneous breaking of the dark $U(1)^\prime$ gauge symmetry, the dark bosons $A^\prime$ and $h_D$ obtain their masses as
$m_{A^\prime}=g_{D} v_{D}$ and $m_{h_{D}}=\sqrt{\lambda_{D}} v_{D}$. 

\begin{figure}[tb]
\centering
\includegraphics[scale=0.3845]{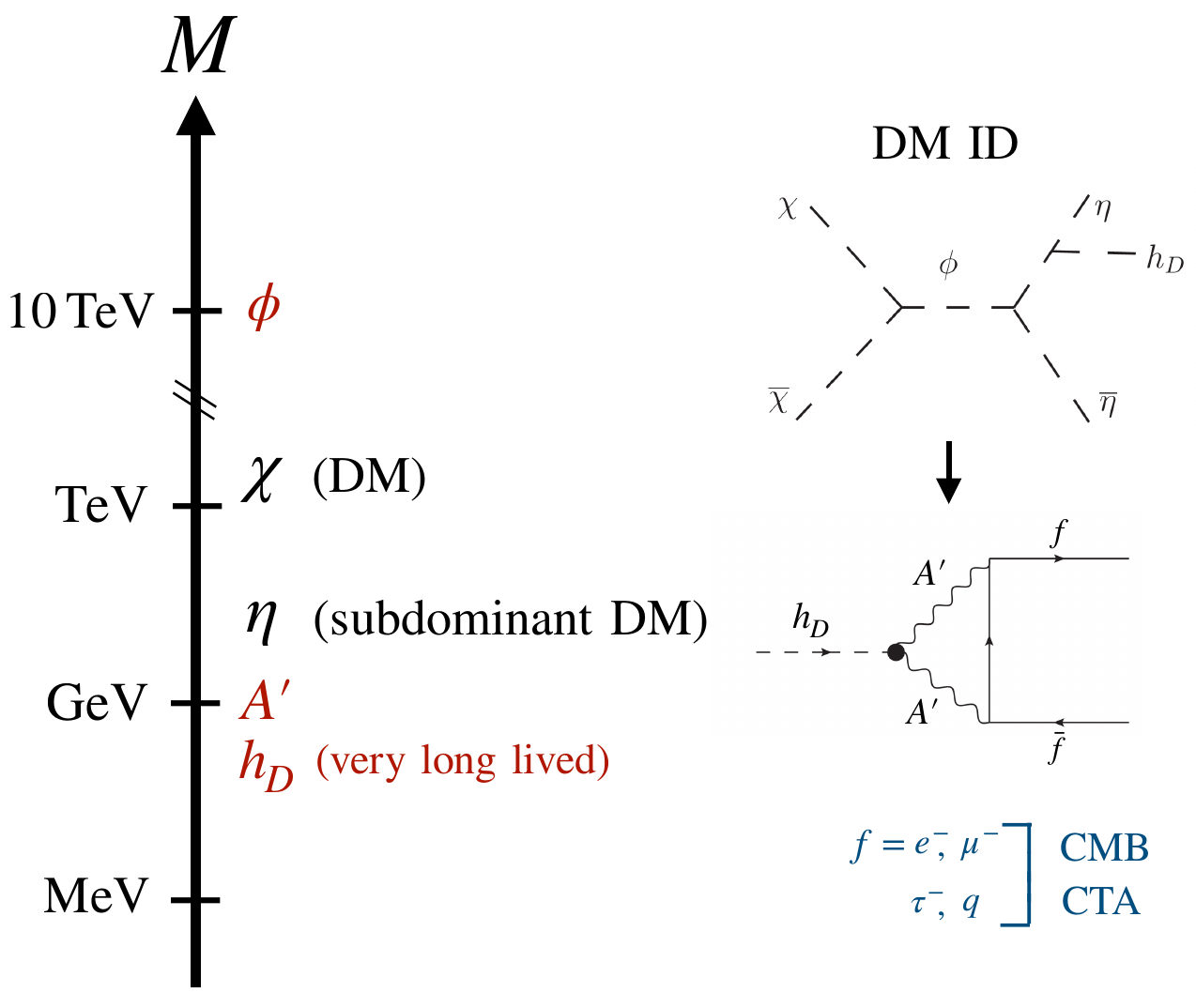}
\caption{
On the left side, a schematic illustration of the mass hierarchy across the energy scales for dark sector particles in the model is shown. The unstable mediators are denoted in dark-red, while the two stable DM species are in black. On the right side, Feynman diagrams for processes leading to DM indirect detection signatures are shown. From top to bottom, these include the $2\to 3$ annihilation process of $\chi$ DM producing the dark Higgs boson $h_D$ and the loop-induced $h_D$ decay into the SM species.
\label{fig:idea}}
\end{figure}

The dark sector spans several orders of magnitude in mass, as shown on the left side of \cref{fig:idea}. Even for all the dark charges set to unity, the model is still characterized by $9$ free parameters. In order to organize our discussion and better highlight interesting phenomenological prospects of this scenario, we assume below the following mass hierarchy in the BSM sector of the model
\begin{equation}
(m_{\phi}\gg)\ m_{\chi} > m_{\eta}>m_{A^\prime}\gtrsim m_{h_D}>2m_{f}.
\label{eq:mass_scheme}
\end{equation}
With this assumption, we find below that the dominant DM component is the heavier scalar field $\chi$ which decouples earlier from thermal plasma than the lighter species $\eta$. In particular, the decoupling of $\chi$ proceeds via annihilations through an exchange of the intermediate heavy scalar, $\chi\chi\to (\phi^\ast)\to \eta\eta$. In addition, $2\to 3$ annihilation processes are possible with an outgoing dark scalar emitted from the $\eta$ leg, $\chi\chi\to (\phi^\ast)\to \eta\eta h_D$. The latter process will play a crucial role in our discussion of ID observables. On the right side of \cref{fig:idea}, we show the respective Feynman diagrams for the key processes leading to indirect signals of DM in our model.

As can be seen, for $m_{h_D}\lesssim m_{A^\prime}$,\footnote{This mass hierarchy corresponds to $\lambda_D \lesssim g_D^2/2$, which can be expected from the renormalization group equation for $\lambda_D$, when one imposes the perturbativity constraint up to $O(\tev)$ scale; see, \eg, Refs~\cite{Batell:2009yf,Darme:2017glc}. Interestingly, such mass hierarchy also excludes the decay of $h_D$ into $A^\prime$ and SM states, which naturally leads to long-lived nature of $h_D$.} the dark Higgs boson produced in the $2\to 3$ process subsequently decays into SM fermions via a triangle loop involving off-shell dark vectors; see \cref{eq:mass_scheme}.\footnote{We note that the $h_D$ decays into 4 charged SM states mediated by two off-shell dark photons, $h_D \to A^{\prime*} A^{\prime*} \to 2 f^+ 2 f^-$, are also possible, albeit they remain subdominant~\cite{Batell:2009yf}.} As a result, the decay width is naturally suppressed, and $h_D$ may have a very large lifetime of astrophysical relevance,
\begin{equation}
c\tau_{h_D} \simeq 1~\textrm{kpc} \left(\frac{0.04}{g_D}\right)^2 \left(\frac{10^{-6}}{\epsilon}\right)^4 \left(\frac{2 \gev}{m_{h_D}}\right) \left(\frac{m_{A^\prime}}{2 \gev}\right)^2\ .
\label{eq:ctauhD}
\end{equation}
The approximate result in \cref{eq:ctauhD} is derived from a full expression given in \cref{eq:DH_SMSM}. As can be seen from \cref{eq:ctauhD}, the dark scalar $h_D$ can travel galactic-scale distances before decaying and producing visible signals, $h_D\to f\bar{f}$. The decay final-states are $f\bar{f} = e^+e^-$, $\mu^+\mu^-$, $\tau^+\tau^-$, or hadrons formed due to hadronization of the light quark pairs $q\bar{q}$. Instead, we require $A^\prime$ to decay before the Big Bang Nucleosynthesis (BBN) in the early Universe~\cite{Berger:2016vxi}, \ie, $\tau_{A^\prime}\lesssim 0.1\second$. In fact, the lifetime of the dark photon is often within the reach of the intensity frontier searches for light new physics. We provide a complete list of decay width and annihilation cross sections relevant to our discussion in \cref{sec:sec_formulae}.

%%%%%%%%%%%%%%%%%%%%%%%%%%%%%%%%%%%%%%%%%%%%%%%%%%%%
\section{Relic density\label{sec:relic}}
%%%%%%%%%%%%%%%%%%%%%%%%%%%%%%%%%%%%%%%%%%%%%%%%%%%%

\paragraph{Boltzmann equations} We begin the discussion of specific features of our model by examining the relic density of both DM components, the dominant one $\chi$ and a minor one $\eta$ -- for which we require $\Omega_{\chi}h^2+\Omega_{\eta}h^2\simeq 0.12$~\cite{Planck:2018vyg} -- as well the abundance of the unstable long-lived dark Higgs $h_D$. To this end, we numerically solve a set of Boltzmann equations that extend the familiar assisted freeze-out mechanism discussed for two-component dark sectors~\cite{Belanger:2011ww} to compute the relic densities of three dark species:\footnote{We solve the Boltzmann equations assuming a partial wave decomposition of the thermally averaged annihilation cross sections for each process, $\langle \sigma v\rangle$. Due to large mass hierarchies among the three species, the times of their thermal freeze-out are well separated, which means that it is not essential to keep their full thermal dependence of $\langle \sigma v\rangle$.}
\begin{align}
\label{eq:boltzmann_three}
\frac{d Y_{\chi}}{d x} &=-\frac{\lambda_{\chi}}{x^{2}}\left(Y_{\chi}^{2}-\frac{Y_{\eta}^{2}}{\left(Y_{\eta}^{\mathrm{eq}}\right)^{2}} \left(Y_{\chi}^{\mathrm{eq}}\right)^{2} \right) \\
& - \frac{\tilde{\lambda}_{\chi}}{x^{2}}\left(Y_{\chi}^{2}-\left(Y_{\chi}^{\mathrm{eq}}\right)^{2} \frac{Y_{\eta}^2}{\left(Y_{\eta}^{\mathrm{eq}}\right)^2} \frac{Y_{h_D}}{\left(Y_{h_D}^{\mathrm{eq}}\right)} \right), \\
\frac{d Y_{\eta}}{d x} &=-\frac{\lambda_{\eta}}{x^{2}}\left(Y_{\eta}^{2}-\left(Y_{\eta}^{\mathrm{eq}}\right)^{2}\frac{Y_{h_D}^{2}}{\left(Y_{h_D}^{\mathrm{eq}}\right)^{2}}\right) \\
&+ \frac{\lambda_{\chi}}{x^{2}}\left(Y_{\chi}^{2}-\frac{Y_{\eta}^{2}}{\left(Y_{\eta}^{\mathrm{eq}}\right)^{2}} \left(Y_{\chi}^{\mathrm{eq}}\right)^{2} \right) \nonumber\\
& + \frac{2}{3}\frac{\tilde{\lambda}_{\chi}}{x^{2}}\left(Y_{\chi}^{2}-\left(Y_{\chi}^{\mathrm{eq}}\right)^{2} \frac{Y_{\eta}^2}{\left(Y_{\eta}^{\mathrm{eq}}\right)^2} \frac{Y_{h_D}}{\left(Y_{h_D}^{\mathrm{eq}}\right)} \right),\nonumber\\
\frac{d Y_{h_D}}{d x} &= \frac{\lambda_{\eta}}{x^{2}}\left(Y_{\eta}^{2}-\left(Y_{\eta}^{\mathrm{eq}}\right)^{2}\frac{Y_{h_D}^{2}}{\left(Y_{h_D}^{\mathrm{eq}}\right)^{2}}\right) \\
&-\frac{\lambda_{h_D}}{x^{2}}\left(Y_{h_D}^{2}-\left(Y_{h_D}^{\mathrm{eq}}\right)^{2} \right) \nonumber\\
&+ \frac{1}{3}\frac{\tilde{\lambda}_{\chi}}{x^{2}}\left(Y_{\chi}^{2}-\left(Y_{\chi}^{\mathrm{eq}}\right)^{2} \frac{Y_{\eta}^2}{\left(Y_{\eta}^{\mathrm{eq}}\right)^2} \frac{Y_{h_D}}{\left(Y_{h_D}^{\mathrm{eq}}\right)} \right),\nonumber
\end{align}
where $Y_i$ is the respective yield, $i = \chi, \eta, h_D$, and $x = m_{\eta}/T$. We stress again that we assume the lifetime of the dark photon to be small such that $A^\prime$ decays before the era of BBN. Therefore, further analysis does not require a precise determination of its abundance. We assume, however, that $A^\prime$ remains in thermal contact with the SM radiation around the time of $h_D$ freeze-out. In fact, we will focus below on scenarios with a low mass splitting between these two dark species,\footnote{We note that for the small mass splitting $\Delta_{h_D A^\prime}\ll 1$, the coannihilation process $A^\prime h_D\to (A^{\prime\ast})\to f\bar{f}$ is also possible, where $f$ is the SM fermion. This could further suppress the primordial relic abundance of the dark Higgs boson. However, the small kinetic mixing parameter $\epsilon$ suppresses the relevant cross section and, therefore, this process does not significantly affect $\Omega_{h_D}$ in our analysis.}
\begin{equation}
\Delta_{h_D A^\prime} = \frac{m_{A^\prime} - m_{h_D}}{m_{A^\prime}}\ll 1.
\label{eq:masssplitting}
\end{equation}
The parameters $\lambda_i$ depend on the annihilation cross sections of the processes that play the dominant role in determining the abundances in the dark sector. Given mass ordering shown in \cref{eq:mass_scheme}, the abundance of the main DM component depends on
\begin{equation}
{\lambda_{\chi} \equiv \frac{s(m_{\eta})}{H(m_{\eta})}\left\langle\sigma_{\chi \bar{\chi} \rightarrow \eta \bar{\eta}} v\right\rangle \simeq \frac{1.32\,g_{*s}(m_{\eta})}{\sqrt{g_{*}(m_{\eta})}}m_{\eta} m_{\textrm{Pl}} \left\langle\sigma_{\chi \bar{\chi} \rightarrow \eta \bar{\eta}} v\right\rangle},
\end{equation}
while $\lambda$'s for the other leading processes are
\begin{align*}
&\lambda_{\eta} \equiv \frac{s(m_{\eta}) }{H(m_{\eta})}\left\langle\sigma_{\eta \bar{\eta} \rightarrow h_D h_D} v\right\rangle,\\
&\lambda_{h_D} \equiv \frac{s(m_{\eta}) }{H(m_{\eta})}\left\langle\sigma_{h_D h_D \rightarrow A^\prime A^\prime} v\right\rangle, \\
&\tilde{\lambda}_{\chi} \equiv \frac{s(m_{\eta})}{H(m_{\eta})}\left\langle\sigma_{\chi \bar{\chi} \rightarrow \eta \bar{\eta} h_D} v\right\rangle,
\end{align*}
where $m_{\textrm{Pl}}=2.44 \times 10^{18}\gev$ is the reduced Planck mass, $s\equiv s(T)$ is the entropy density and $H\equiv H(T)$ is the Hubble rate. The effective number of degrees of freedom for the entropy and energy densities of the thermalized SM-DS plasma at temperature $T$ is denoted by $g_{*s}(T)$ and $g_{*}(T)$, respectively. The equilibrium comoving yield of each particle is defined as $Y^{\mathrm{eq}}_i=n^{\mathrm{eq}}_i/s$, and it explicitly reads:
\begin{equation}
Y_{i}^{\mathrm{eq}}(x)=\frac{g_i}{g_{*s}(x)} \frac{45}{4 \pi^{4}} (r_i\,x)^{2} K_{2}[r_i\,x].
\end{equation}
Here, $r_i=m_i/m_{\eta}$ and the number of internal degrees of freedom of particle $i$ is denoted by $g_i$. The resulting relic density is obtained from $\Omega_{i} h^2=(\rho_{i}/\rho_{\mathrm{crit}})h^2=(s_0\,Y_{i,0}\,m_{i}/\rho_{\mathrm{crit}})h^2$, where $s_0$ is the present-day entropy density and $Y_{i,0}$ is the final yield of the dark species $i$ after its freeze-out.

In all cases that we consider, $\chi$ is the dominant DM component, $\Omega_{\chi}h^2\simeq 0.12$, which undergoes ordinary freeze-out almost independently of the other species because it is much heavier than them; see \cref{eq:mass_scheme}.
The values of the heavy (auxiliary) scalar $\phi$ mass and the coupling constants $\mu_{\chi}$ and $\mu_{\eta}$, see \cref{eq:SASA_SBSB}, are chosen such that the total $\chi \chi$ annihilation cross section is equal to the thermal value $\left\langle\sigma v\right\rangle_{\chi\chi \to \eta\eta} + \left\langle\sigma v\right\rangle_{\chi\chi \to \eta\eta h_D} = 2.2\times 10^{-26}\, \mathrm{cm}^3/\mathrm{s}$. 

The abundances of lighter species - $\eta$ and $h_D$ - depend on the couplings $\lambda_{h_D \eta}$ and $g_D$, respectively, see \cref{eq:SBSB_DHDH} and \cref{eq:DPDP_DHDH}. These couplings must be large, $\lambda_{h_D \eta}$, $g_D \gtrsim 0.1$. This is because we require the relic abundance of $h_D$ to be suppressed to avoid stringent cosmological constraints; see \cref{sec:constraintsandfuture}, and we are interested in regions of the parameter space predicting a large value of the cross section relevant to DM indirect detection, $\chi\chi \to \eta\eta h_D$.
As a result, we typically find in our analysis
\begin{equation}
    \Omega_{\chi}h^2\simeq 0.12 \gg \Omega_{\eta}h^2 \gtrsim \Omega_{h_D}h^2\ .
\end{equation}
Our discussion assumes that kinetic equilibrium is established until chemical decoupling. As discussed in~\cite{Pospelov:2007mp}, this can be obtained for sufficiently large values of the kinetic mixing parameter
\begin{equation}
    \epsilon \gtrsim 10^{-6} \times \left(\frac{m_{\chi}}{500 \gev}\right)^2 \left(\frac{10 \gev}{m_{A^\prime}}\right) .
\end{equation}    
This also guarantees that $A^\prime$ decays early, before the era of BBN.

\paragraph{Balanced and standard freeze-out regimes} In a large part of the allowed parameter space of the model, the lighter species $\eta$ and the long-lived $h_D$, undergo the so-called balanced freeze-out~\cite{Agashe:2014yua}, which results in phenomenological features different from the ordinary freeze-out mechanism. 
In this regime, the relic density of the sub-dominant DM component $\eta$ or a long-lived dark scalar $h_D$ is characterized by the following scaling $\Omega\propto 1/\sqrt{\left\langle\sigma v\right\rangle}$, instead of the usual scaling $\Omega\propto 1/\left\langle\sigma v\right\rangle$ characteristic for standard freeze-out.
This can be obtained when the annihilation cross sections of the sub-dominant DM components are orders of magnitude greater than the annihilation cross section of the dominant DM component. In the balanced freeze-out case, we predict the following simple scaling between the relic densities of the dominant and sub-dominant components
\begin{equation}
    \Omega_{\eta} \simeq \Omega_{\chi} \sqrt{\frac{\lambda_{\chi}+\frac23\tilde{\lambda}_{\chi}}{\lambda_{\eta}}}\frac{m_\eta}{m_\chi},
    \label{eq:balanced_eta_2}
\end{equation}
which can be formally obtained by setting $dY_\eta/dx=0$ in \cref{eq:boltzmann_three}. We note that this relation takes into account the possible production of $\eta$s in annihilations of $\chi$ into both two- and three-body final states, $\chi\chi\to\eta\eta(h_D)$. A similar relation holds for the relic abundance of the $h_D$ species, $\Omega_{h_D} \simeq \Omega_{\chi} [(\lambda_{\chi}+\tilde{\lambda}_{\chi})/\lambda_{h_D}]^{1/2}\,(m_{h_D}/m_\chi)$.
From this expression, it is also clear that the balanced freeze-out regime for $h_D$ points towards large values of $g_D \gtrsim 0.1$, as $\lambda_{h_D}\sim g_D^4$ in \cref{eq:DPDP_DHDH}, and small dark Higgs boson mass $m_{h_D}/m_\chi \ll 1$. This helps to suppress the relevant relic density. The same holds for $\eta$ when $\lambda_{h_D \eta} \gtrsim 0.5$.

\begin{figure}[tb]
\centering
\includegraphics[scale=0.93]{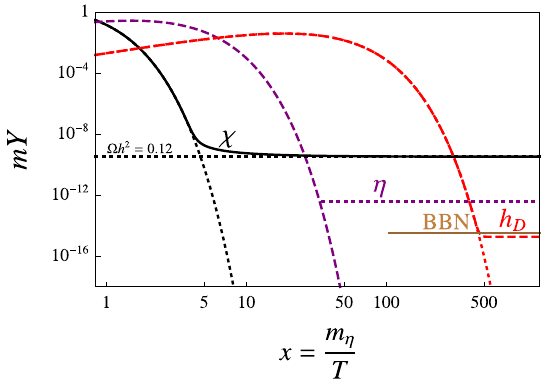}
\caption{
Comoving energy densities, $(mY)_i$, for two-component DM and the unstable dark Higgs boson undergoing thermal freeze-out obtained by numerically solving the set of Boltzmann equations given by \cref{eq:boltzmann_three}. We fixed the dark sector masses at the following values: $m_{\chi}=1.5~\tev$, $m_{\eta}=250~\gev$, $m_{h_D}=3~\gev$, and the mass splitting $\Delta_{h_D A^\prime} = 0.01$; see \cref{eq:masssplitting}. We also fixed the coupling constants $g_D=0.4$ and $\lambda_{h_D \eta}=0.5$, while the value of $\mu_{\chi}$, $\mu_{\eta}$, and $m_\phi$ are chosen such that $\langle\sigma v\rangle_{\chi\chi\to \eta\eta}$ is at thermal value and $\Omega_{\chi}h^2\simeq 0.12$, as indicated with the horizontal black dotted line. The brown horizontal line denotes the BBN limit on late-time decaying $h_D$ with $\tau_{h_D}\lesssim 10^{12}~\textrm{s}$. \label{fig:relic_density_three_component_benchmarks}
}
\end{figure}

By solving the relevant Boltzmann equations, we have numerically verified that the aforementioned simple expressions hold in appropriate regions of the model's parameter space and for benchmark scenarios presented below. We illustrate the evolution of the yield $m_i\,Y_i$ for all the three dark species as a function of $x$ in the right panel of \cref{fig:relic_density_three_component_benchmarks}. We assume the following mass benchmark: $m_{\chi} = 1.5~\tev$, $m_{\eta} = 250~\gev$, $m_{h_D} = 3~\gev$, and a small mass splitting between the dark Higgs boson and dark photon $\Delta_{h_D A^\prime} = 0.01$. As discussed in \cref{sec:model}, we keep the dark vector mass larger than the scalar mass to avoid direct decays of $h_D$ into the $A^\prime$ species. At the same time, for small mass splittings, the relic density of $h_D$ in the early Universe is still determined by the annihilation process $h_Dh_D\to A^\prime A^\prime$ in the forbidden DM regime~\cite{DAgnolo:2015ujb}. In the figure, we also fix the relevant coupling constants $g_D=0.4$ and $\lambda_{h_D \eta}=0.5$. As one can see, for sufficiently large $g_D$, the $h_D$ abundance may become strongly suppressed with respect to the total DM relic density. In particular, for the assumed values of the model parameters shown in the figure, we obtain $\Omega_{h_D}\lesssim 10^{-5}\,\Omega_{\chi}$. With such a small abundance, one can avoid stringent BBN bounds for even very long-lived dark Higgs bosons. We illustrate relevant BBN constraint in the plot assuming $\tau_{h_D}\lesssim 10^{12}~\textrm{s}$.

%%%%%%%%%%%%%%%%%%%%%%%%%%%%%%%%%%%%%%%%%%%%%%%%%%%%%%%%%%%
\section{Current and future constraints from astrophysics, cosmology and colliders\label{sec:constraintsandfuture}}
%%%%%%%%%%%%%%%%%%%%%%%%%%%%%%%%%%%%%%%%%%%%%%%%%%%%%%%%%%%

Although most of the dark sector species in the model considered here remain secluded from the SM, their indirect couplings via a light dark vector portal $A^\prime$ still induce experimental constraints. Below, we briefly summarize the current constraints on our scenario. We also discuss future experimental and observational prospects.

\subsection{Accelerator-based searches\label{sec:accelerators}}

Light secluded dark photons are among the primary targets of intensity frontier searches for sub-$\gev$ new physics as they correspond to one out of only a few available simple renormalizable portals to the dark sector. Based on available data, one can put upper bounds on the kinetic mixing parameter $\epsilon$. Of particular relevance to our study are constraints obtained from CHARM~\cite{CHARM:1985anb,Gninenko:2012eq}, E141~\cite{Riordan:1987aw,Andreas:2012mt}, NuCal~\cite{Blumlein:2013cua,Blumlein:1991xh}, Orsay~\cite{Andreas:2012mt,Davier:1989wz} experiments, and the newest bounds obtained by the FASER~\cite{CERNFASER} and NA62~\cite{NA62:2023qyn,CERNNA62} collaborations.

As discussed above, rare meson decays provide stringent bounds on light, sub-$\gev$ dark vector bosons from various past accelerator-based experiments. Similar searches in the future are expected to constrain further the available parameter space of the model. For the kinetic mixing parameter range of our interest, important bounds are expected to come from the ongoing and proposed LLP searches in beam-dump and collider experiments; see Refs~\cite{Beacham:2019nyx,Alimena:2019zri,Antel:2023hkf} for reviews. In the following, we will show selected such bounds from the DarkQuest~\cite{Apyan:2022tsd}, FASER2~\cite{Feng:2017uoz,FASER:2018eoc}, LHCb~\cite{Craik:2022riw}, NA62~\cite{NA62:2023qyn}, and SHiP~\cite{Alekhin:2015byh} experiments.

We note that direct searches for heavier dark species at the LHC will be very challenging since we assume that they couple to the SM only via the $A^\prime$ portal and the corresponding kinetic mixing parameter is suppressed. Similarly, the dark Higgs boson in our setup remains only very weakly coupled to the SM, and it avoids accelerator-based constraints.

\subsection{Astrophysical and cosmological bounds\label{sec:CMB}} 

Additional constraints on light dark vectors can be obtained from their possible impact on astrophysical and cosmological observations. The dark photon of our interest, however, avoids bounds due to possible modifications of the supernovae cooling rate and the neutrino emission from SN1987A~\cite{Chang:2018rso}, and BBN constraints on late-time decaying unstable BSM species~\cite{Berger:2016vxi}.

Important  bounds arise due to the residual relic abundance of potentially very long-lived dark Higgs $h_D$. Following Ref.~\cite{Poulin:2016anj}, we implement relevant BBN constraints on late-time electromagnetic energy injection. We also use the results of Ref.~\cite{Lucca:2019rxf} for the CMB bounds based on the combined data from the Planck~\cite{Planck:2018vyg} and COBE/FIRAS~\cite{Fixsen:1996nj} satellite missions. In the latter case, we follow Refs~\cite{Slatyer:2015jla,Slatyer:2015kla} and modify the CMB constraints from Ref.~\cite{Lucca:2019rxf} by taking into account a finite fraction of the electromagnetic energy transferred to the intergalactic medium, $f_{\textrm{eff}}<1$, characterizing different SM final states present in the model. The final shape of the cosmological bounds derived this way depends on the interplay between the $h_D$ lifetime; see \cref{eq:ctauhD}, and its relic abundance obtained by solving the Boltzmann equations discussed in \cref{sec:relic}.

Future surveys are expected to improve the bounds on CMB spectral distortions significantly and to essentially exclude BSM scenarios predicting the mediator lifetime between $\tau_{h_D}\sim 10^{5}\second$ and $10^{12}\second$~\cite{Lucca:2019rxf}. The relevant bounds will constrain the relic abundance of unstable, long-lived species to be not larger than a tiny fraction of the total DM relic density, $\Omega_{h_D}\lesssim 10^{-6}\,\Omega_{\textrm{DM}}$, but, depending on $\tau_{h_D}$, even much more stringent limits can be derived of order $10^{-12}\,\Omega_{\textrm{DM}}$. Instead, for the CMB anisotropy data, the expected improvement over current bounds is less pronounced but will also result in more stringent constraints by about a factor of a few in $\Omega_{h_D}$. This is relevant for the large lifetime regime, $\tau_{h_D} > 10^{12}\second$.

In the following, we will present expected combined bounds from the Planck~\cite{Planck:2018vyg} and the proposed Primordial Inflation Explorer (PIXIE)~\cite{Kogut:2011xw} satellite mission. In addition, we will also employ more stringent constraints expected from the combination of the future data from the Polarized Radiation Imaging and Spectroscopy Mission (PRISM)~\cite{PRISM:2013fvg}, ground-based CMB Stage-4 (CMB S-4) searches~\cite{CMB-S4:2016ple} and the space mission LiteBIRD~\cite{Matsumura:2013aja}. We stress that, while the CMB data remain complementary to DM ID searches for mediator lifetimes $\tau_{h_D}\gtrsim 10^{5}\second$, they will provide the best way of probing scenarios with extremely long-lived $h_D$s that would predict much-suppressed ID signal rates.

We note that both the current and future CMB bounds can be additionally affected by the impact of late-time DM annihilations, $\chi\chi\to\eta\eta h_D$, producing a subdominant abundance of very long-lived dark Higgs bosons in between the $\chi$ DM freeze-out and the CMB epoch. The typical energy of such $h_D$s results in the relevant boost factors of the order $\gamma_{h_D}\sim 10$ to $100$ for the benchmark scenarios considered below. Hence, such subdominant $2\to 3$ DM annihilation processes occurring even around the matter-radiation equality could produce light Higgs bosons that are not completely redshifted before the recombination period. The impact of the CMB bounds and future searches could then be modified in the regime of the large $h_D$ lifetime close to the current bounds, e.g., $\tau_{h_D}\gtrsim 10^{11}~\textrm{s}$ in the plots shown in \cref{sec:results}. While this effect is not expected to affect our discussion below significantly, we leave a relevant detailed analysis for future studies.

\subsection{Dark matter indirect detection\label{sec:DMID}}

The model of our interest can also be probed in DM ID experiments. The corresponding discovery prospects rely on annihilation rates that depend only on unsuppressed couplings in the secluded dark sector. On the other hand, future DD searches remain much less promising. The lighter DM species $\eta$ couple to the SM only indirectly. Their suppressed couplings to light quarks and negligible abundance typically yield very low expected signal rates in DD experiments. This is also true for the heavier DM  species $\chi$ for which the scattering rates off nuclei appear only via intermediate $\eta$ and $\phi$ fields.

ID signatures in our model can arise from annihilations of both DM species, the heavier dominant scalar $\chi$ and the lighter one $\eta$. The latter process, however, is highly suppressed by the tiny abundance of $\eta$. This is not the case for annihilations of the dominant DM component. At the leading order, though, the main annihilation channel is purely into invisible final states $\chi\chi\to \eta\eta$. Instead, the dominant contribution to ID signal rates appears at the next-to-leading order due to the $2\to 3$ process $\chi\chi\to \eta\eta h_D$ shown in \cref{fig:idea}. This is especially important for a growing value of the dark coupling constant $\lambda_{h_D\eta}$ which increases the probability of emitting the final-state $h_D$ off the $\eta$ leg; see \cref{eq:MSASA}.\footnote{The dark Higgs bosons could also be produced in cross interactions $\chi\eta\to \chi\eta h_D$ which, however, play a subdominant role with respect to the $2\to 3$ annihilations of the $\chi$s due to the suppressed relic density of $\eta$.} Notably, as we have argued above, larger values of $\lambda_{h_D\eta}$ are also preferred by cosmological and collider constraints.

We note that DM ID prospects in our model rely on the $2\to 3$ annihilation process producing a continuous (not monochromatic) spectrum of the meta-stable $h_D$ mediator energies. This results in smearing the spectrum of final-state SM particles; see \cref{app:spectrum} for discussion about the photon spectrum. Hence, one effectively avoids bounds from the searches for the peaked spectral features in the positron data; see Refs~\cite{Bergstrom:2013jra,John:2021ugy}. On top of this, interesting non-local effects can further affect DM ID event rates in this scenario~\cite{Rothstein:2009pm,Kim:2017qaw,Chu:2017vao,Gori:2018lem,Agashe:2020luo}. We discuss them in more detail in \cref{app:nonlocal}.

A particularly promising way of probing these scenarios is via searches for a diffuse DM-induced $\gamma$-ray flux. The corresponding differential flux of photons coming to a detector from the angular region in the sky $\Delta\Omega$ is given by
\begin{equation}
\left(\frac{d\Phi}{dE_\gamma}\right)_{\textrm{standard}} = \frac{1}{8\pi}\frac{\langle\sigma v\rangle}{m_{\chi}^2}\int_{\Delta\Omega}d\Omega\,\int_{\textrm{l.o.s.}}\rho_{\chi}^2\,ds\,\left(\frac{dN_\gamma}{dE_\gamma}\right)_{\chi},
\label{eq:flux}
\end{equation}
where on the left-hand side, we have denoted that this result corresponds to the ``standard'' regime in which the long-lived mediator $h_D$ decays after traveling distances much shorter than a kpc. In our analysis below, we employ the Einasto DM profile,
\begin{equation}
\rho_\chi\equiv\rho_{\text {Einasto}}(r)=\rho_{s} \exp \left(-\frac{2}{\alpha}\left[\left(\frac{r}{r_{s}}\right)^{\alpha}-1\right]\right),
\label{eq:Einasto}
\end{equation}
where $\rho_{s}=0.079\gev/\mathrm{cm}^3$, $r_{s}=20~\mathrm{kpc}$ and $\alpha=0.17$~\cite{Pieri:2009je}.
In \cref{eq:flux} the photon spectrum from $2\to 3$ annihilations of $\chi$ and subsequent decays of $h_D$ is denoted by $(dN_\gamma/dE_\gamma)_{\chi}$; see \cref{app:spectrum} for further discussion. Importantly, the above chain of processes results in a much softer photon spectrum than one could expect based on the DM mass. In particular, a typical $h_D$ energy after the initial $\chi$ annihilation is of the order $E_{h_D}\sim (0.1 - 0.2)\,m_{\chi}$. The peak energy of photons produced as a result of $h_D$ decays is then further shifted towards smaller energies, $E_\gamma\lesssim\textrm{few tens of}~\gev$. However, it contains a high-energy tail extending towards larger $E_\gamma$.

In \cref{sec:results}, we will present the expected sensitivity reach of the forthcoming DM ID experiment targeting heavy DM species, namely the Cherenkov Telescope Array (CTA)~\cite{CTAConsortium:2017dvg}; see also Refs~\cite{Hryczuk:2019nql,CTA:2020qlo}. Instead of performing full detector simulations for the scenario of our interest, which is beyond the scope of our study, we illustrate the impact of the CTA on the parameter space of the model with some approximate expected bounds. We obtain them by employing the CTA sensitivity plots for the secluded DM regime following Ref.~\cite{Siqueira:2021lqj}, which we, however, modify by taking into account characteristic $h_D$ energies obtained after the $2\to 3$ annihilation process. We expect this simplified procedure to encompass the most essential effects of our study correctly. Notably, the applicable CTA bounds for secluded DM only mildly depend on the DM mass in the regime between $100~\gev \lesssim m_{\textrm{DM}}\lesssim \textrm{few}~\tev$. 

%%%%%%%%%%%%%%%%%%%%%%%%%%%%%%%%%%%%%%%%%%%%%%%%%%
\section{Results\label{sec:results}}
%%%%%%%%%%%%%%%%%%%%%%%%%%%%%%%%%%%%%%%%%%%%%%%%%%

The non-minimal content of our model results in $9$ free parameters that contain the masses of the dark species and various dark sector couplings; see \cref{sec:model}. 
Our goal in this paper has been to point out some new signatures that arise in the model, rather than to perform a thorough investigation of its rich phenomenology.
We, therefore, limit our discussion to only slices of this multidimensional parameter space. Below, we first justify our choice for the fixed parameters, and later we present the results of our analysis in a convenient two-dimensional ($m_{A^\prime}$, $g_D$) plane.

As discussed above, we work in the regime of a small mass splitting between the dark Higgs boson and the dark photon, $\Delta_{h_D A^\prime} = 0.01$ in \cref{eq:masssplitting}. We employ three different values of the kinetic mixing parameter $\epsilon = 10^{-4}$, $10^{-5}$, and $10^{-6}$. We set $m_{\chi} = 1.5\tev$ for the heavy DM particle and $m_{\eta} = 250\gev$ for the lighter one. In our plots, the $\eta$ DM component is kept heavier than the dark Higgs boson, $m_{h_D} < m_\eta$. This allows for efficient annihilations $\eta\eta\to h_Dh_D$ and suppresses the $\eta$ abundance to negligible levels. Otherwise, it will significantly contribute to the total DM relic density and could be overproduced. We fit the auxiliary parameters in the dark sector, $m_\phi$, $\mu_\chi$, and $\mu_\eta$, such that at each point in the parameter space shown in the figures below, the heavy scalar DM obtains the correct value of the thermal relic density, $\Omega_{\chi}h^2\simeq 0.12$. In contrast, $\Omega_{\eta}h^2$ is negligible.

The values of the dark coupling constant $g_D$ are only limited by the perturbativity bound. We take it to be $\alpha_D/4\pi<1$, which originates from the fact that such combination enters into 1-loop corrections. This condition is equivalent to $g_D< 4\pi$. The astrophysical and cosmological bounds discussed in \cref{sec:constraintsandfuture} constrain this coupling constant from below.
In particular, lower values of $g_D$ would lead to a too large abundance of $h_D$, and its late-time decays would violate BBN and CMB limits. Instead, collider bounds affect the region of the parameter space with too light dark photons below a few hundred MeV. All the bounds are shown as gray-shaded regions in the figures.

\begin{figure*}[tb]
    \centering
    \includegraphics[scale=0.384]{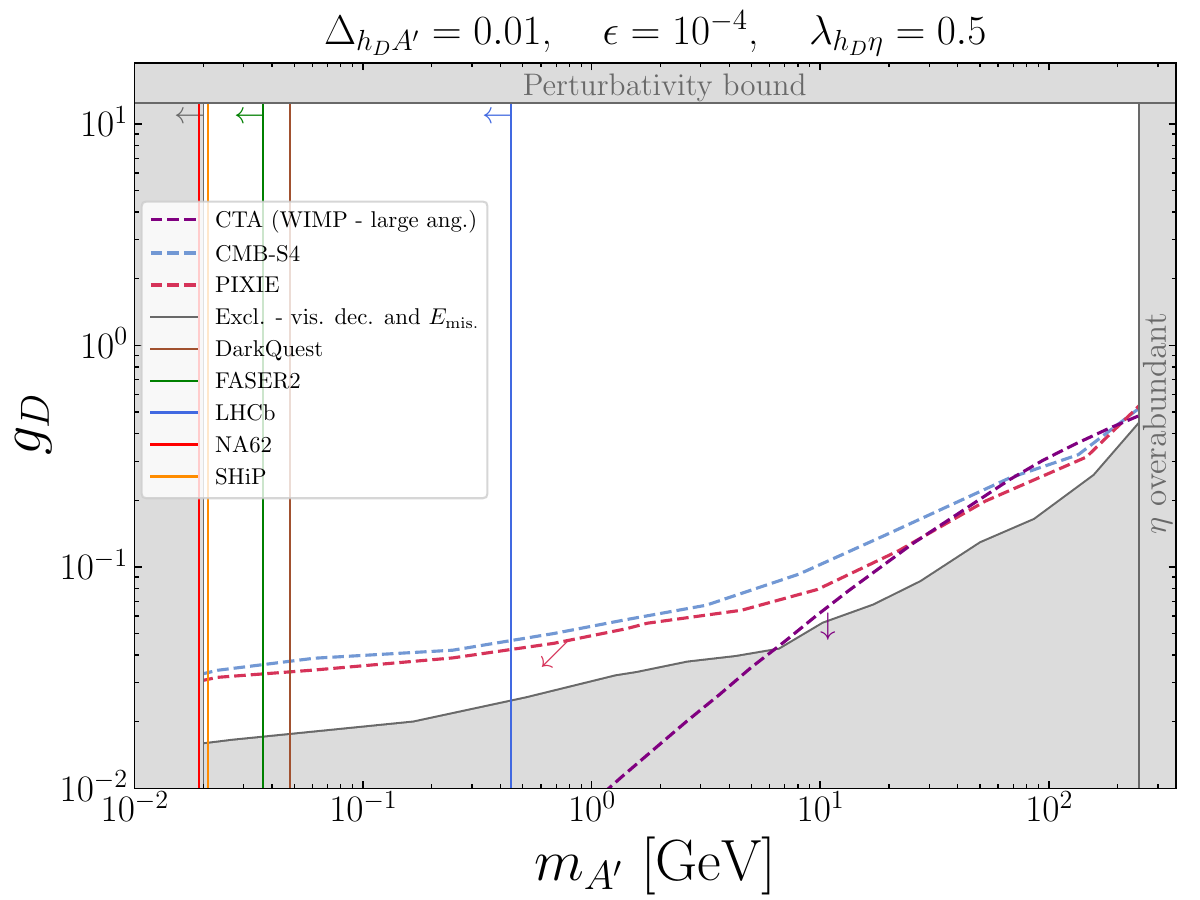}\hspace*{0.22cm}
    \includegraphics[scale=0.384]{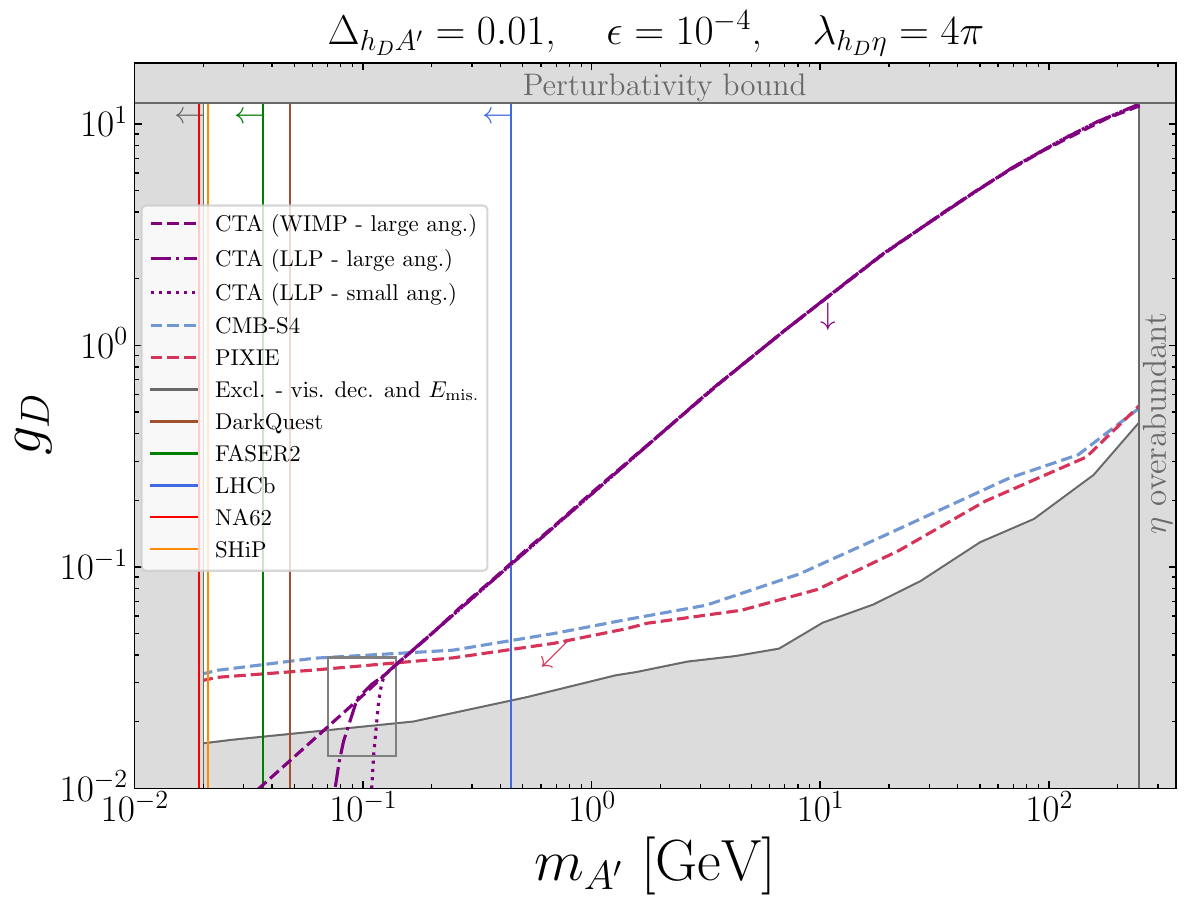}\vspace*{0.3cm}
    \includegraphics[scale=0.384]{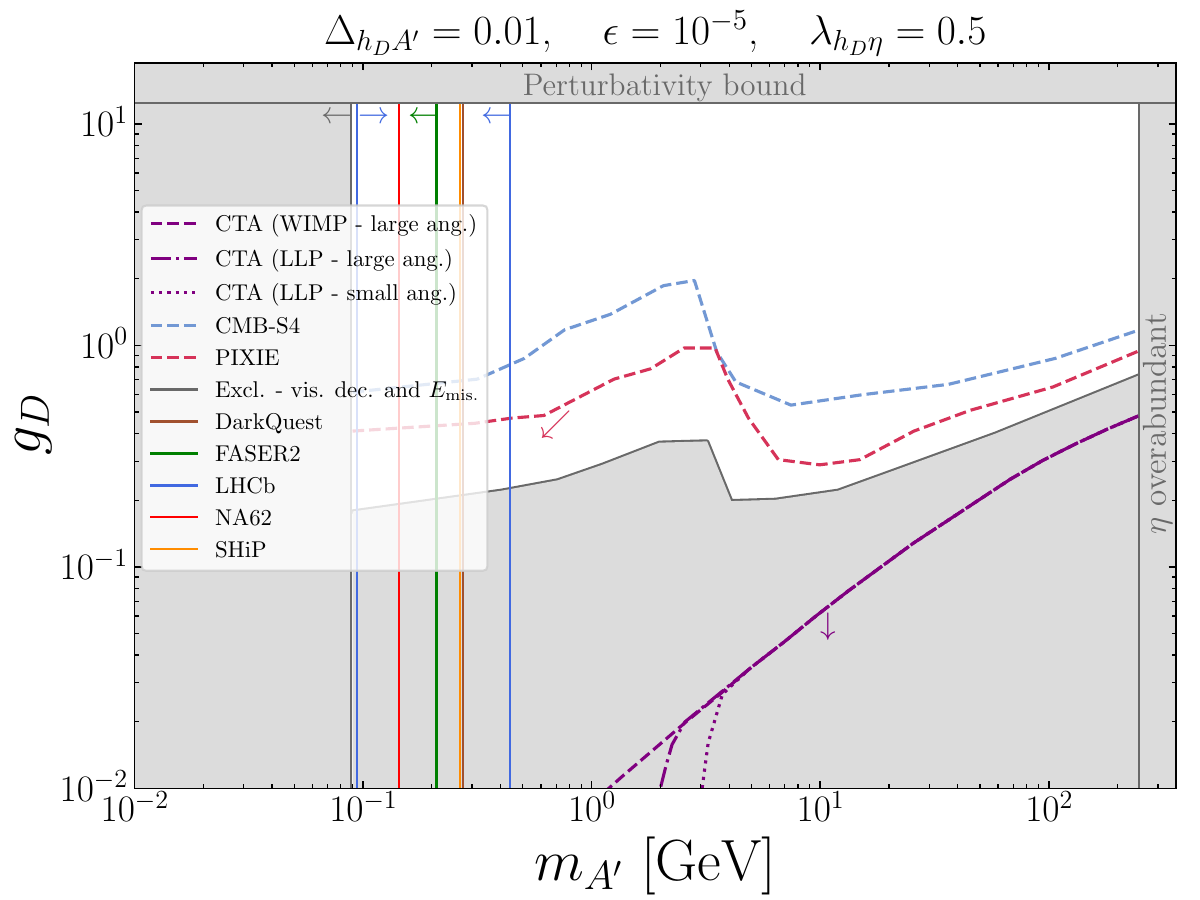}\hspace*{0.22cm}
    \includegraphics[scale=0.384]{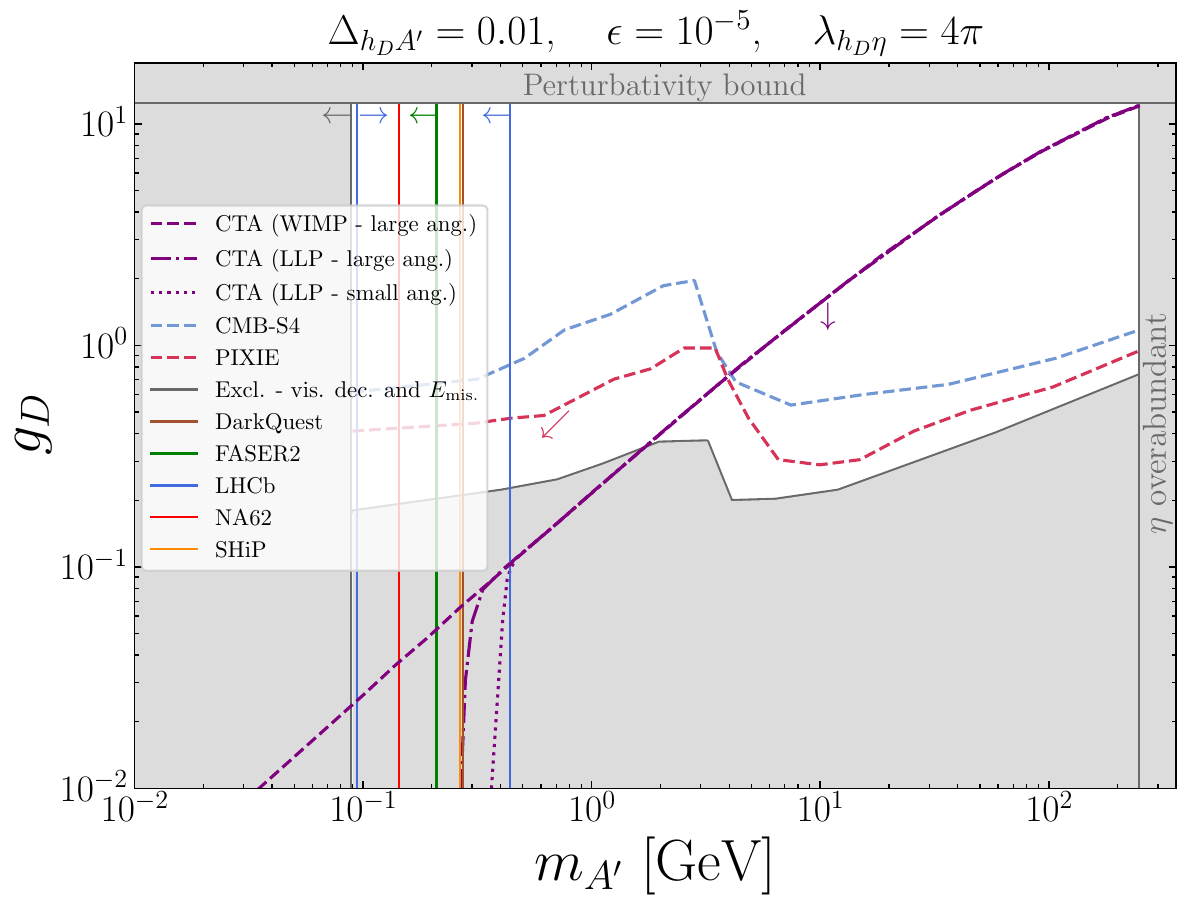}\vspace*{0.3cm}
    \includegraphics[scale=0.384]{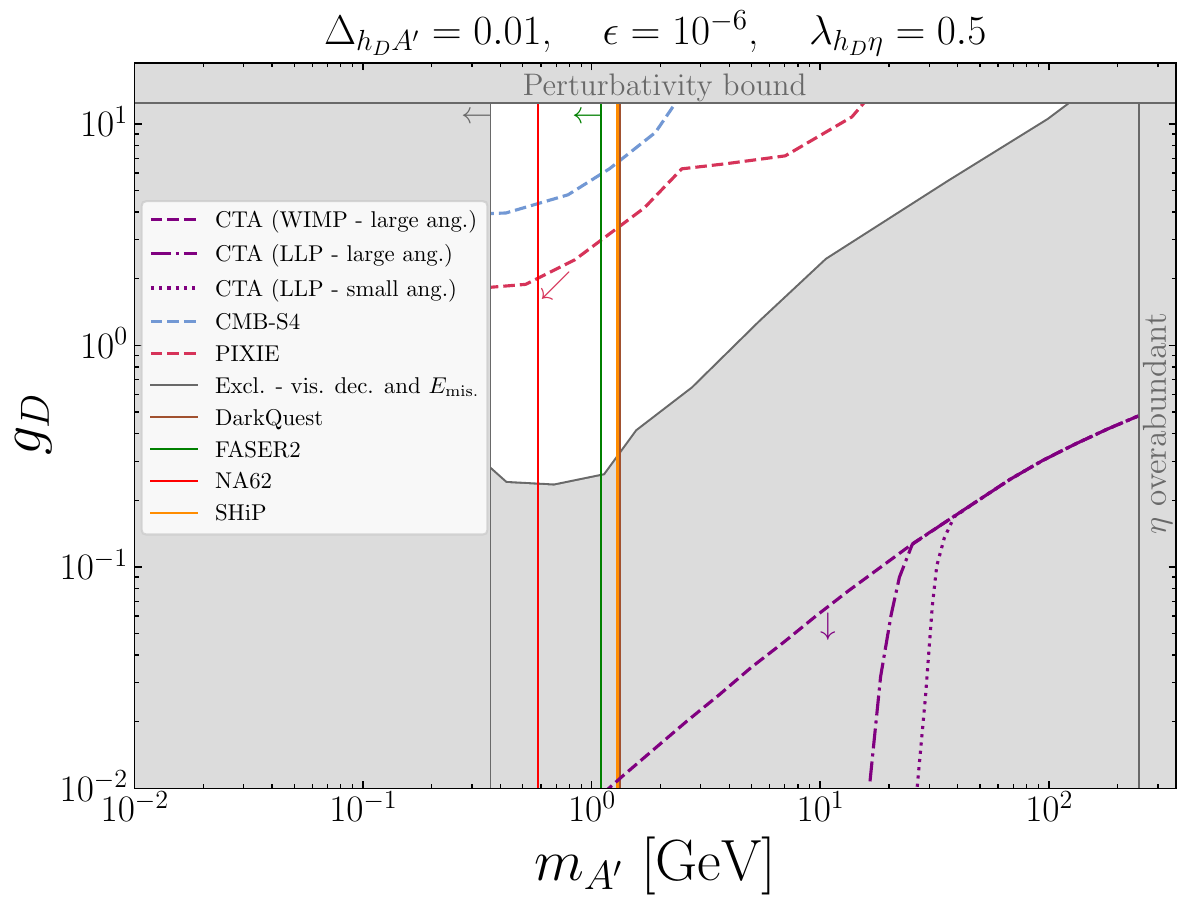}\hspace*{0.22cm}
    \includegraphics[scale=0.384]{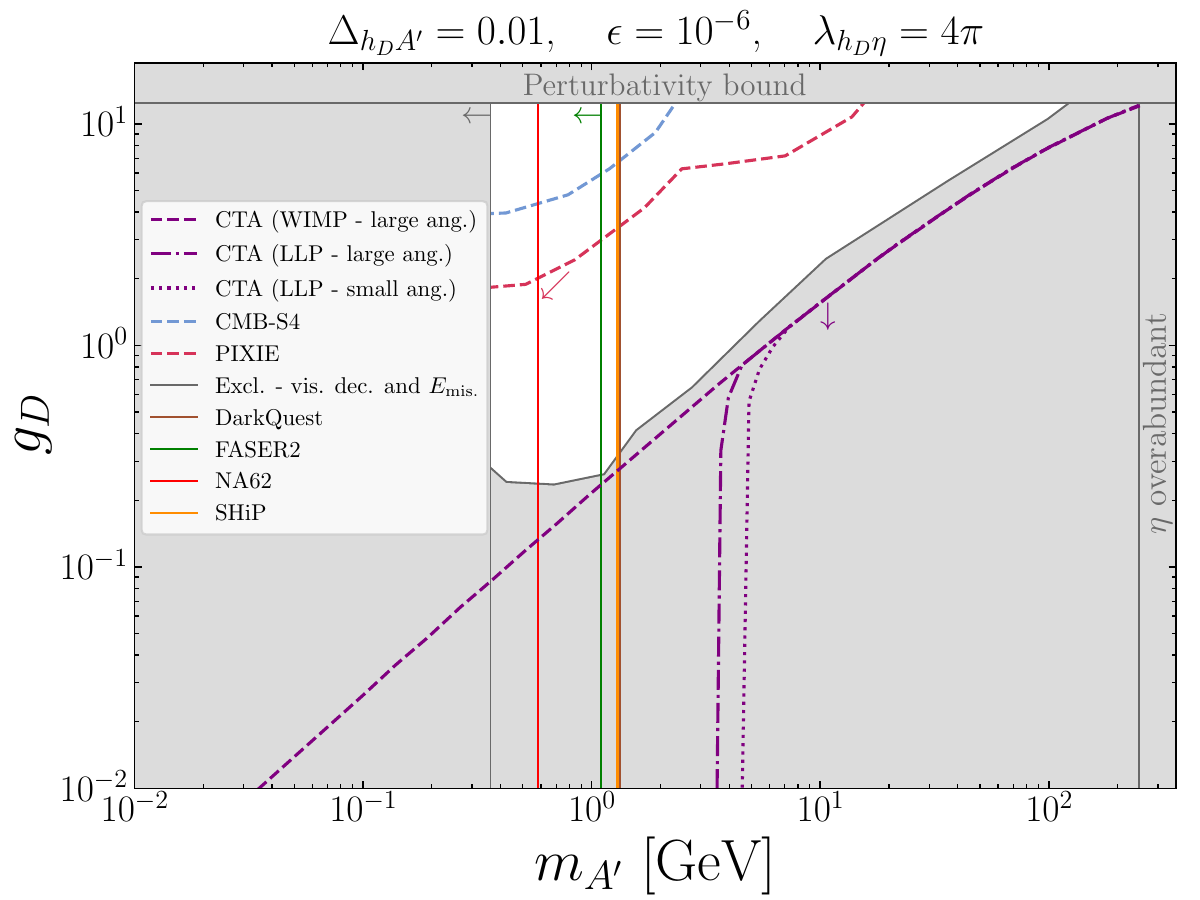}
    \caption{
    The impact on the parameter space of the considered model of various past bounds (gray-shaded regions) and future searches is shown in the $(m_{A^\prime},g_D)$ plane; see the text for details. We fix other parameters of the model to $m_{\chi} = 1.5~\tev$, $m_{\eta} = 250~\gev$, and $m_{h_D} = 0.99 \times m_{A^\prime}$. We also fix the kinetic mixing parameter to $\epsilon = 10^{-4}$ ($10^{-5}$, $10^{-6}$) in the top (central, bottom) panels, and we assume $\lambda_{h_D \eta}=0.5$ ($4\pi$) in the left (right) panels. In each point in the figure, the parameters $m_\phi$, $\mu_{\chi}$, and $\mu_{\eta}$ are chosen such that the correct value of the heavy DM relic density is obtained, $\Omega_{\chi}h^2\simeq 0.12$. 
    \label{fig:results_mDM_gD_1}
    }
\end{figure*}

We present expected sensitivity reach lines of future searches in \cref{fig:results_mDM_gD_1}. The top (central, bottom) panels correspond to $\epsilon = 10^{-4}$ ($10^{-5}$, $10^{-6}$). On the left, we fix the coupling strength between the dark Higgs boson and $\eta$ DM component to $\lambda_{h_D\eta} = 0.5$. In the right panels, we assume it saturates the perturbativity bound $\lambda_{h_D\eta} = 4\pi$. 
We note that in our analysis, we assumed for simplicity that the bare value of $\mu_{h_D \eta}$ vanishes, and instead $\mu_{h_D \eta}= \lambda_{h_D\eta} v_D$ is obtained only after the $U(1)_{\mathrm{dark}}$ spontaneous symmetry breaking. A non-zero bare value of $\mu_{h_D \eta}$ could lower the value of $\lambda_{h_D\eta}$ corresponding to the regime of long-lived $h_D$ significantly below $\sim O(1)$. 
As can be seen, in each case, the allowed region of the parameter space corresponds to the $A^\prime$ mass in between tens or hundreds MeV and $m_{A^\prime}\simeq m_\eta = 250~\textrm{GeV}$. The dark coupling constant varies in this region between $g_D\sim (0.01-0.1)$ and the perturbativity bound.

Scenarios with the sub-GeV dark photon can be constrained in the future in collider-based searches, as we show with vertical, colorful lines in the plots. These projected constraints and past bounds do not depend on the coupling strengths $g_D$ and $\lambda_{h_D\eta}$, while they vary for different values of the kinetic mixing parameter $\epsilon$. Future LHCb searches are expected to provide the most severe such bounds for $\epsilon = 10^{-4}$ and $10^{-5}$. For $\epsilon = 10^{-6}$, the dominant bounds will be found by DarkQuest, FASER2, and SHiP experiments.

Complementary bounds from future CMB surveys are indicated in the plots with light blue and red dashed lines. They will exclude too low values of $g_D$, for which the $h_D$ lifetime grows; see \cref{eq:ctauhD}. Instead, these means do not constrain the region in the model's parameter space with large values of $g_D$, close to the perturbativity limit. The relic abundance of $h_D$ remains, in this case, highly suppressed beyond the expected sensitivity of the future CMB and BBN analyses. The projected constraints and past bounds become more severe for decreasing $\epsilon$, which is due to its impact on the $h_D$ lifetime in a loop-induced decay process; see \cref{eq:ctauhD}. In particular, for $\epsilon = 10^{-6}$, future CMB-S4 measurements will be able to probe almost the entire remaining available parameter space of this model.

We also show in the plots regions of the parameter space that the future DM ID searches in CTA can probe. These also constrain the available parameter space from below, i.e., they will exclude too low values of $g_D$. This is because the $2\to 3$ annihilation cross section determining the DM ID signal rates decreases for increasing value of this dark coupling constant; see \cref{eq:MSASA} in which $\sigma_{\chi\chi\to\eta\eta h_D}\propto \lambda_{h_D\eta}/g_D$ and we fix $\lambda_{h_D\eta}$ in the plots. We note, however, that for natural values of the dark coupling constant $g_D\lesssim 1$, some regions of the available parameter space of the model might be probed in future DM ID searches. This especially concerns scenarios with increasing $m_{A^\prime}\simeq m_{h_D}$. 
Instead, the $2\to 3$ annihilation cross section is suppressed for decreasing $m_{A^\prime}$. The projected CTA bounds will only mildly constrain the available parameter space of the model for $\lambda_{h_D\eta} = 0.5$ or $\epsilon\lesssim 10^{-5}$. Instead, they become more severe with a growing value of these coupling strengths and become the dominant indirect bounds for $\epsilon = 10^{-4}$ and $\lambda_{h_D\eta} = 4\pi$. The dependence on $\lambda_{h_D\eta}$ is driven by the aforementioned $2\to 3$ annihilation cross section. On the other hand, the DM ID signal rates do not depend on $\epsilon$. However, with decreasing value of this parameter, these bounds become less competitive with CMB and BBN constraints.

For sufficiently low values of $g_D$, the lifetime of the dark Higgs boson is very large, and non-local effects in DM ID become important. These can suppress DM ID signal rates, especially for small regions of interest around the Galactic Center. We show it in the plots with dashed, dotted, and dash-dotted purple lines; see \cref{app:nonlocal} for detailed discussion.

%%%%%%%%%%%%%%%%%%%%%%%%%%%%%%%%%%%%%%%%%%%%%%%%%%%%%
\section{Conclusions\label{sec:conclusions}}
%%%%%%%%%%%%%%%%%%%%%%%%%%%%%%%%%%%%%%%%%%%%%%%%%%%%%

Light sub-$\gev$ portal models of dark matter have gained much interest in recent years and have sparked both experimental and theoretical activity in the field. Most studies so far have focused on simplified frameworks with only a limited number of new species added to extend the SM. These typically correspond to the most popular interaction operators and are supposed to encompass essential phenomenological aspects of such simplified scenarios. While this approach allows one to do a manageable comparison among many experimental proposals, it is also understood that new effects might be observed in more elaborate, and more realistic, models incorporating a larger number of BSM species with a mass hierarchy that can be quite complex and in particular span several orders of magnitude.
In this paper we have presented an example of such a scenario.

The presence of the light portal particle that thermally connects the SM sector and a heavy DM in the early Universe leaves important observational imprints that can be distinct from those in both more popular heavy WIMP models and from scenarios predicting the existence of only light DM. In particular, new detection prospects can appear, in this case, due to the interplay between intensity frontier searches, cosmological observations, and DM indirect detection.

In this study, we have found such effects and exposed them in a model which extends a popular scenario with a dark photon portal to DM and employs the dark Higgs boson vev to drive the light dark sector masses. We have discussed the consequences of adding a new heavy DM particle to this setup coupled predominantly to the lighter DM particle, ensuring its secluded nature differs from traditional WIMPs. This scenario remains beyond the reach of current and near-future direct detection experiments and collider searches for heavy DM. However, we have shown here that the best way of probing this model type, besides intensity frontier searches for LLPs, is to employ DM-induced signatures in future ID and CMB experiments.

The presence of unstable subdominant dark species produced in annihilations of much heavier dominant DM particles can lead to further striking signatures. We have illustrated this in the case of the dark Higgs boson $h_D$ that can feature a very large loop-suppressed lifetime and an astrophysically interesting decay length $\gamma\beta c\tau_{h_D}\sim 1\,\textrm{kpc}$. This can lead to interesting non-local effects in DM ID searches, for instance, enhanced DM signal rates from the GC and simultaneously suppressed corresponding rates expected from the dwarf galaxies. A thorough experimental testing of such scenarios will require going beyond the traditional approach to DM ID.

In searching for new physics, it remains essential to encompass a broad range of possible BSM scenarios and to study all potential complementarities among different types of experiments. An attractive framework based on the popular thermal DM paradigm has led to numerous studies in the past years. Models of this class that predict the existence of a light sub-$\gev$ portal to heavy secluded DM deserve special attention in such efforts. As shown here, their characteristic signatures are absent in more popular simplified scenarios but may be successfully probed in extensive experimental programs in the coming years.

\paragraph{Acknowledgements} We would like to thank Brian Batell for useful remarks and comments on the manuscript. We would like to thank Luc Darm$\acute{\textrm{e}}$, Iftah Galon, Andrzej Hryczuk, Arvind Rajaraman for useful discussions. In our analysis, we employed python module \texttt{vegas}~\cite{peter_lepage_2021_4746454}, which uses an algorithm developed in Refs~\cite{Lepage:1977sw,Lepage:2020tgj}. KJ and LR are supported by the National Science Centre, Poland, research grant No. 2015/18/A/ST2/00748. LR and ST are supported by the grant ``AstroCeNT: Particle Astrophysics Science and Technology Centre'' carried out within the International Research Agendas programme of the Foundation for Polish Science financed by the European Union under the European Regional Development Fund. ST is supported by the National Science Centre, Poland, research grant No. 2021/42/E/ST2/00031. KJ is supported by the IBS under the project code, IBS-R018-D1.

\appendix

%%%%%%%%%%%%%%%%%%%%%%%%%%%%%%%%%%%%%%%%%%%%%%%%%%%%%%%%%%%%%
\section{Particle physics formulae\label{sec:sec_formulae}}
%%%%%%%%%%%%%%%%%%%%%%%%%%%%%%%%%%%%%%%%%%%%%%%%%%%%%%%%%%%%%

Below, we provide expressions for all the decay widths and scattering cross sections relevant for our analysis.

\subsection{Decay widths}
\paragraph{Decay width of $A^\prime$} 
It is given by
\begin{equation}
    \Gamma_{A^\prime} = \frac{\Gamma_{A^\prime\rightarrow e^+e^-}}{\textrm{B}_e(m_{A^\prime})},
\end{equation}
where
\begin{equation*}
    \Gamma_{A^\prime\rightarrow e^+e^-} = \frac{\epsilon^2\,e^2\,m_{A^\prime}}{12\,\pi}
    \! \times \!
    \bigg[1-\frac{4 m_{e^-}^2}{m_{A^\prime}^2}\bigg]^{\frac 12}\!\!\!
    \! \times \!
    \bigg[1+\frac{2\,m_{e^-}^2}{m^2_{A^\prime}}\bigg],
    \label{eq:AprimeGamma}
\end{equation*}
where $m_{e^-}$ is the electron mass and $\textrm{B}_e = \textrm{B}(A^\prime\to e^+e^-)$ is the branching fraction of a decay into an electron-positron pair, which can be found, e.g., in Ref.~\cite{Bauer:2018onh}.

\paragraph{Decay width of $h_D$}
The dark Higgs plays the role of the LLP in our setup, which can be naturally achieved due to loop suppression of the decay width into SM fermions~\cite{Batell:2009yf}:
\begin{align}
\label{eq:DH_SMSM}
\Gamma_{h_D \to e^+e^-} &= \frac{\alpha_D \alpha^2 \varepsilon^4 m_{h_D}}{2 \pi^2} \frac{m_{e^-}^2}{m_{A^\prime}^2}\left(1-4 \frac{m_{e^-}^2}{m_{h_D}^2}\right)^{3 / 2} \\
& \times \left|I\left(\frac{m_{h_D}^2}{m_{A^\prime}^2}, \frac{m_{e^-}^2}{m_{A^\prime}^2}\right)\right|^2, \nonumber
\end{align}
where the $I$ loop function is given by
\begin{equation}
I\left(x_s, x_e\right)=\int_0^1 d y \int_0^{1-y} d z \frac{2-(y+z)}{(y+z)+(1-y-z)^2 x_e-y z x_s},
\end{equation}
and an analogous expression holds for $h_D$ decays into pairs of other SM charged states.

\paragraph{Decay width of $\phi$} While it is not essential for our study, we also provide for completeness the decay width of the heavy scalar $\phi$, which can decay into both the $\chi\bar{\chi}$ and $\eta\bar{\eta}$ pairs:
\begin{equation}
\Gamma_\phi=\Gamma_{\phi\to \chi \chi}+\Gamma_{\phi\to \eta \eta}=\frac{\mu_{\chi}^2\sqrt{m_\phi^2-4m_{\chi}^2}}{16\pi m_\phi^2}+\frac{\mu_{\eta}^2\sqrt{m_\phi^2-4m_{\eta}^2}}{16\pi m_\phi^2}\ .
\end{equation}
In the assumed mass scheme, \cref{eq:mass_scheme}, and for $\mu_{\chi}=\mu_{\eta}$, $\mu_{\chi}/m_\phi \sim 0.1$, we obtain a very small lifetime of $\phi$, which then does not have any implications on our discussion of the cosmological bounds
\begin{equation}
\tau_\phi \sim \frac{8\pi m_\phi}{\mu_{\chi}^2} \sim 10^{-25}\second\left(\frac{1~\tev}{\mu_{\chi}}\right)^2 \frac{m_\phi}{10~\tev}\ .
\label{eq:tau_phi}
\end{equation}

\subsection{Annihilation cross sections}
\begin{figure}[tb]
\centering
\includegraphics[width=0.23\textwidth]{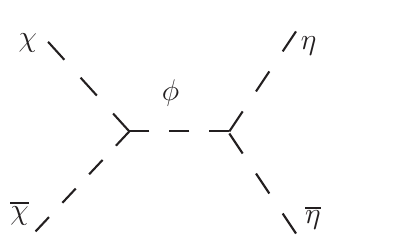}
\includegraphics[width=0.5\textwidth]{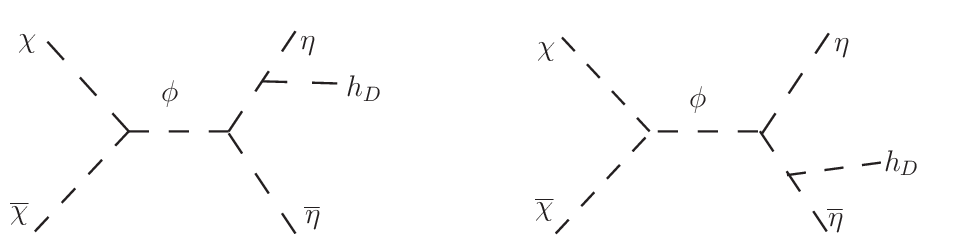}
\caption{Heavy DM annihilation processes contributing to the $\chi$ relic density.}
\label{fig:CMB_SASA}
\end{figure}

\paragraph{$\chi \chi \to \eta \eta$}
The relic density of $\chi$ is driven by annihilations into two lighter species $\eta$, as shown on the left in \cref{fig:CMB_SASA}. The relevant cross section is given by
\begin{align}
\label{eq:SASA_SBSB}
\left\langle \sigma v \right\rangle_{\chi\chi\to \eta\eta} &= \frac{\mu_{\chi}^2\mu_{\eta}^2 \sqrt{m_{\chi}^2-m_{\eta}^2}}{32\pi m_{\chi}^3\left[(m_\phi^2-4m_{\chi}^2)^2+\Gamma_\phi^2 m_\phi^2\right]} \nonumber\\
& \sim 10^{-9}\,\gev^{-2} \left(\frac{\frac{\mu}{m_\phi}}{0.1}\right)^4 \left(\frac{30\gev}{m_{\chi}}\right)^2\ .
\end{align}

\begin{figure}[tb]
    \centering
    \includegraphics[width=0.15\textwidth]{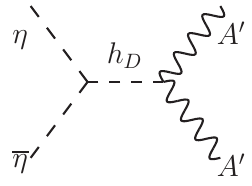}
    \caption{Feynman diagram for the $\eta$ annihilation processes into a pair of dark vectors.}
    \label{fig:CMB_SBSB_DPDP}
\end{figure}
    
\begin{figure}[tb]
\centering
\includegraphics[width=0.5\textwidth]{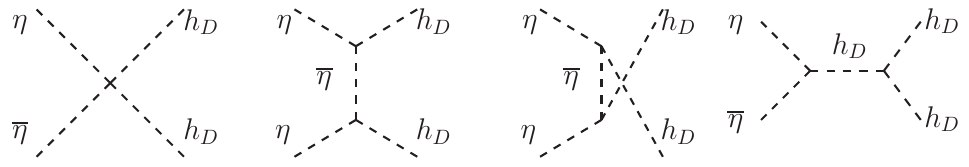}
\caption{Feynman diagrams for the $\eta$ annihilation processes into a pair of dark Higgs boson.}
\label{fig:CMB_SBSB_DHDH}
\end{figure}

\begin{figure}[tb]
    \centering
    \includegraphics[width=0.5\textwidth]{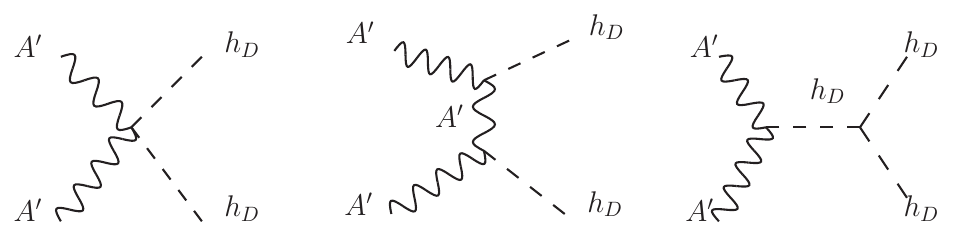}
    \caption{Feynman diagrams for the dark vector annihilations into light dark Higgs bosons.}
    \label{fig:CMB_DPDP_DHDH}
\end{figure}

\paragraph{$\eta\eta\to A^\prime A^\prime$} 
The relic density of $\eta$ is obtained via its annihilations into the light mediators, $A^\prime$ and $h_D$, as shown in \cref{fig:CMB_SBSB_DPDP,fig:CMB_SBSB_DHDH}, and are described by the following formula:
\begin{align}
\label{eq:SBSB_DPDP}
\left\langle \sigma v \right\rangle_{\eta\eta\to A^\prime A^\prime} = & \frac{\lambda_{h_D \eta}^2 \sqrt{m_{\eta}^2-m_{A^\prime}^2}}{16 \pi m_{\eta}^3 (m_{h_D}^2 - 4 m_{\eta}^2)^2} \nonumber\\
&(4 m_{\eta}^4 + 3 m_{A^\prime}^4 - 4 m_{\eta}^2 m_{A^\prime}^2) ,
\end{align}
which, in the limit of $m_{\eta} \gg m_{A^\prime}, m_{h_D}$, simplifies to
\begin{equation}
\left\langle \sigma v \right\rangle_{\eta\eta \to A^\prime A^\prime} = \frac{\lambda_{h_D \eta}^2}{64 \pi m_{\eta}^2}\ .
\label{eq:SBSBApAp}
\end{equation}

\paragraph{$\eta\eta\to h_D h_D$} The cross section for $\eta$ annihilations into light dark Higgs bosons reads
\begin{align}
    \left\langle \sigma v \right\rangle_{\eta\eta\to h_D h_D} = & \frac{\lambda_{h_D \eta}^2 \sqrt{m_{\eta}^2-m_{h_D}^2}}{64 \pi  g_D^4 m_{\eta}^3 \left(m_{h_D}^4-6 m_{h_D}^2 m_{\eta}^2+8 m_{\eta}^4\right)^2} \nonumber \\ 
    & \times \big[g_D^2 \left(5 m_{h_D}^4-18 m_{h_D}^2 m_{\eta}^2 + 16 m_{\eta}^4\right) \\
    &-2 \lambda_{h_D \eta} m_{A^\prime}^2 \left(m_{h_D}^2-4 m_{\eta}^2\right)\big]^2 \nonumber\ ,
\end{align}
which, in the limit of $m_{\eta} \gg m_{h_D}$, simplifies to
\begin{equation}
    \left\langle \sigma v \right\rangle_{\eta\eta\to h_D h_D} = \frac{\lambda_{h_D \eta}^2}{16 \pi m_{\eta}^2}\ .
\label{eq:SBSB_DHDH}
\end{equation}

\paragraph{$h_D h_D \to A^\prime A^\prime$} The metastable relic density of the dark Higgs is obtained thanks to its annihilations into $A^\prime A^\prime$ pairs, as shown in \cref{fig:CMB_DPDP_DHDH}. As we are working in the forbidden DM regime, $m_{h_D}<m_{A^\prime}$, we give the formula for the cross section for the opposite reaction, $A^\prime A^\prime \to h_D h_D$:
\begin{equation}
    \left\langle \sigma v\right\rangle_{A^\prime A^\prime \rightarrow h_D h_D} = \frac{2 g_{\mathrm{D}}^{4} \sqrt{1-r}\left(r^2-2r-2\right)^2}{9\pi m_{A^\prime}^{2}\left(8-6 r+r^{2}\right)^{2}} \sim 0.01\frac{g_{\mathrm{D}}^{4}}{m_{A^\prime}^{2}}\ ,
\label{eq:DPDP_DHDH}
\end{equation}
where $r=m_{h_D}^2/m_{A^\prime}^2\ll 1$. 
We connect the two processes by using the detailed balance condition, $\left\langle \sigma v\right\rangle_{h_D h_D \rightarrow A^\prime A^\prime} = \left\langle \sigma v\right\rangle_{A^\prime A^\prime \rightarrow h_D h_D} \times \left(n^{\mathrm{eq}}_{A^\prime}/n^{\mathrm{eq}}_{h_D}\right)^2$.

\paragraph{$\chi \chi \to \eta \eta h_D$} For small values of the dark gauge coupling $g_D$, the dark bremsstrahlung process becomes important, because the coupling between a pair of $\eta$s and a $h_D$ is proportional to the dark Higgs vev, $v_D=m_{A^\prime}/g_D$.
The amplitude of the process depends on, i.a., the momenta of all the particles and the total energy in the CM frame $s=(p_1+p_2)^2$:
\begin{align}
\label{eq:MSASA}
M_{\chi \chi \to \eta \eta h_D} &= \frac{\mu_{\chi} \mu_{\eta}}{(s-m_\phi^2)} \frac{\lambda_{h_D \eta} m_{A^\prime}}{g_D} \nonumber\\
& \times \left(\frac 1{(p_3+p_4)^2 - m_\eta^2} + \frac 1{(p_3+p_5)^2 - m_\eta^2} \right),
\end{align}
where the incoming $\chi$ particles have momenta $p_1$, $p_2$, the outgoing $\eta$ particles have momenta $p_4$, $p_5$ and the outgoing $h_D$ has momentum $p_3$. The total cross section is given by\footnote{We provide a general expression valid for five different masses, expanding the expressions from~\cite{romaoTechniquesCalculationsQuantum}.}
\begin{align}
\label{eq:brehmstr}
    &\sigma_{\chi \chi \to \eta \eta h_D} = \frac{1}{512 \pi^{4} \sqrt{\lambda(s,m_1^2,m_2^2)}} \\
    & \times\int_{E_{3}^{\min}}^{E_{3}^{\max}} d E_{3} \sqrt{E_{3}^2-m_3^2} \int_{0}^{\pi} d \theta \sin \theta \int_{0}^{\pi} d \theta^{*} \sin \theta^{*}\int_{0}^{2 \pi} d \varphi^{*} \nonumber \\ 
    & \frac{\sqrt{((m_{45}+m_4)^2-m_5^2)((m_{45}-m_4)^2-m_5^2)}}{m_{45}^2} |\overline{M_{\chi \chi \to \eta \eta h_D}}|^{2}.\nonumber
\end{align}
To obtain this expression, we have already performed a trivial integral over the azimuth angle $\phi$, which gives a factor of $2\pi$. The angles written without the asterisk are evaluated in the CM frame of a pair of particles with momenta $p_1$ and $p_2$. In turn, quantities with the asterisk ($^{*}$) are evaluated in the CM frame of pair of particles with momenta $p_4$ and $p_5$. For example, $\theta$ is the scattering angle between the particle with momentum $p_3$ and the collision axis of the particles with momenta $p_1$ and $p_2$ in their CM frame, while $\theta^{*}$ and $\phi^{*}$ denote the angles of the particle with momentum $p_3$ in the frame where $\vec{p}_4+\vec{p}_5=0$.

In our case, we set:
\begin{equation}
\begin{aligned}
m_1 &=m_2=m_{\chi}, \\
m_3 &= m_{h_D},\quad m_4 =m_5=m_{\eta}, \\
E_{3}^{\min}&=m_3,\quad E_{3}^{\max}=\frac{s+m_3^2-(m_4+m_5)^2}{2\sqrt{s}}.
\end{aligned}
\end{equation}
In order to obtain the quantities with the asterisk, one needs to use the following boost factor
\begin{equation}
\begin{aligned}
\gamma &=\frac{E_{45}}{m_{45}},\ \  \beta =\sqrt{1-\frac{1}{\gamma^{2}}}.
\end{aligned}
\end{equation}
The invariant mass $m_{45}$ and the energy of the pair $E_{45}$ are given by
\begin{equation}
\begin{aligned}
m_{45} &=\sqrt{\left(p_{4}+p_{5}\right)^{2}}=\sqrt{s-2 \sqrt{s} E_{3}+m_3^2}, \\
E_{45} &=\frac{s+m_{45}^{2}-m_3^2}{2 \sqrt{s}}.
\end{aligned}
\end{equation}
We also write down the expressions for all components of $p_{4}$ and $p_{5}$:
\begin{equation}
\left\{\begin{array}{l}
p_{4}^{* 0}=\frac{m_{45}^{2}+m_{4}^{2}-m_5^2}{2 m_{45}} \\
p_{4}^{* 1}=p_{45}^{\mathrm{CM}} \sin \theta^{*} \cos \varphi^{*} \\
p_{4}^{* 2}=p_{45}^{\mathrm{CM}} \sin \theta^{*} \sin \varphi^{*} \\
p_{4}^{* 3}=p_{45}^{\mathrm{CM}} \cos \theta^{*}
\end{array}, \quad\left\{\begin{array}{l}
p_{5}^{* 0}=\frac{m_{45}^{2}+m_{5}^{2}-m_4^2}{2 m_{45}} \\
p_{5}^{* 1}=-p_{45}^{\mathrm{CM}} \sin \theta^{*} \cos \varphi^{*} \\
p_{5}^{* 2}=-p_{45}^{\mathrm{CM}} \sin \theta^{*} \sin \varphi^{*} \\
p_{5}^{* 3}=-p_{45}^{\mathrm{CM}} \cos \theta^{*}
\end{array}\right.\right.,
\end{equation}
where
\begin{equation}
p_{45}^{\mathrm{CM}}=\sqrt{(p_{4}^{* 0})^2-m_4^2}.
\end{equation}
Finally, we can express $p_4$ and $p_5$ in the CM frame as a function of $(E_3,\theta,\theta^*,\phi^*)$ by applying the following boost and rotation transformations
\begin{equation}
p_{4,5} =\operatorname{Rot}_{\mathbf{y}}(\theta+\pi) \cdot \mathbf{B o o s t}_{\mathbf{z}}(\beta) \cdot p_{4,5}^{*},
\end{equation}
where the usual boost and rotation matrices are given by
\begin{equation}
\text{Boost}_{\mathbf{z}}(\beta)=\left[\begin{array}{cccc}
\gamma & 0 & 0 & \gamma \beta \\
0 & 1 & 0 & 0 \\
0 & 0 & 1 & 0 \\
\gamma \beta & 0 & 0 & \gamma
\end{array}\right], 
\end{equation}
\begin{equation*}
\text{Rot}_{\mathbf{y}}(\theta)=\left[\begin{array}{cccc}
1 & 0 & 0 & 0 \\
0 & \cos \theta & 0 & \sin \theta \\
0 & 0 & 1 & 0 \\
0 & -\sin \theta & 0 & \cos \theta
\end{array}\right].
\end{equation*}

%%%%%%%%%%%%%%%%%%%%%%%%%%%%%%%%%%%%%%%%%%%%%%%%%%%%%%%%%%%%%%%%
\section{DM-induced $\gamma$-ray spectrum\label{app:spectrum}}
%%%%%%%%%%%%%%%%%%%%%%%%%%%%%%%%%%%%%%%%%%%%%%%%%%%%%%%%%%%%%%%%

In this appendix, we provide further details about our modeling of the DM-induced photon spectrum in the model under study. For the primary spectra of $\gamma$ rays, we rely on the PPPC data~\cite{Cirelli:2010xx}; see also Ref.~\cite{Fortin:2009rq} for useful discussion. We denote the relevant spectrum obtained in the rest frame of the dark Higgs boson by $(dN_\gamma/dE_{\gamma}^{\prime})_{h_D}$. We then boost the photon spectrum into the $\chi$ rest frame. 
For each fixed value of the $h_D$ energy, $E_{h_D}$, we obtain the corresponding photon spectrum from the $\chi\chi\to \eta\eta h_D$ process as
\begin{equation}
\left(\frac{d N_\gamma}{d E_\gamma}\right)_{\chi}\Bigg|_{E_{h_D}} \simeq \frac{m_{h_D}}{2\,E_{h_D}}\,\int_{E_\gamma}^{E_{h_D}}{\frac{dE_\gamma^{\prime}}{E_\gamma^{\prime}}\,\left(\frac{dN_\gamma}{dE_{\gamma}^{\prime}}\right)_{h_D}},
\end{equation}
where $E_\gamma$ denotes the photon energy in the $\chi$ rest frame.\footnote{It is, to a good approximation, the Galactic frame of $\chi$.} In order to obtain the final $\gamma$-ray spectrum $(dN_\gamma/dE_\gamma)_{\chi}$, we then convolute this with the actual continuous $h_D$ spectrum from $2\to 3$ annihilations, $\chi\chi\to \eta\eta h_D$; see \cref{eq:brehmstr} for the relevant differential cross section. In order to simplify our analysis, we evaluate this as a discretized weighted average
\begin{equation}
\left(\frac{d N_\gamma}{d E_\gamma}\right)_{\chi} = \sum_{\textrm{bins}\,E_{h_D}}{\frac{\langle\sigma v\rangle_{E_{h_D}}}{\langle\sigma v\rangle}\,\left(\frac{d N_\gamma}{d E_\gamma}\right)_{\chi}\Bigg|_{E_{h_D}}}\ ,
\end{equation}
where $\langle\sigma v\rangle_{E_{h_D}}$ corresponds to the $2\to 3$ annihilation cross section integrated over a limited range of the outgoing $h_D$ energies (within the energy bin centered around $E_{h_D}$), while $\langle\sigma v\rangle$ is the total such cross section. We used 20 equally spaced bins in $x_{h_D} = E_{h_D}/E_{\chi}$ in the logarithmic scale, which has been numerically verified to be sufficient for the results presented in \cref{sec:results}.

\begin{figure}[tb]
\centering
\includegraphics[scale=0.4]{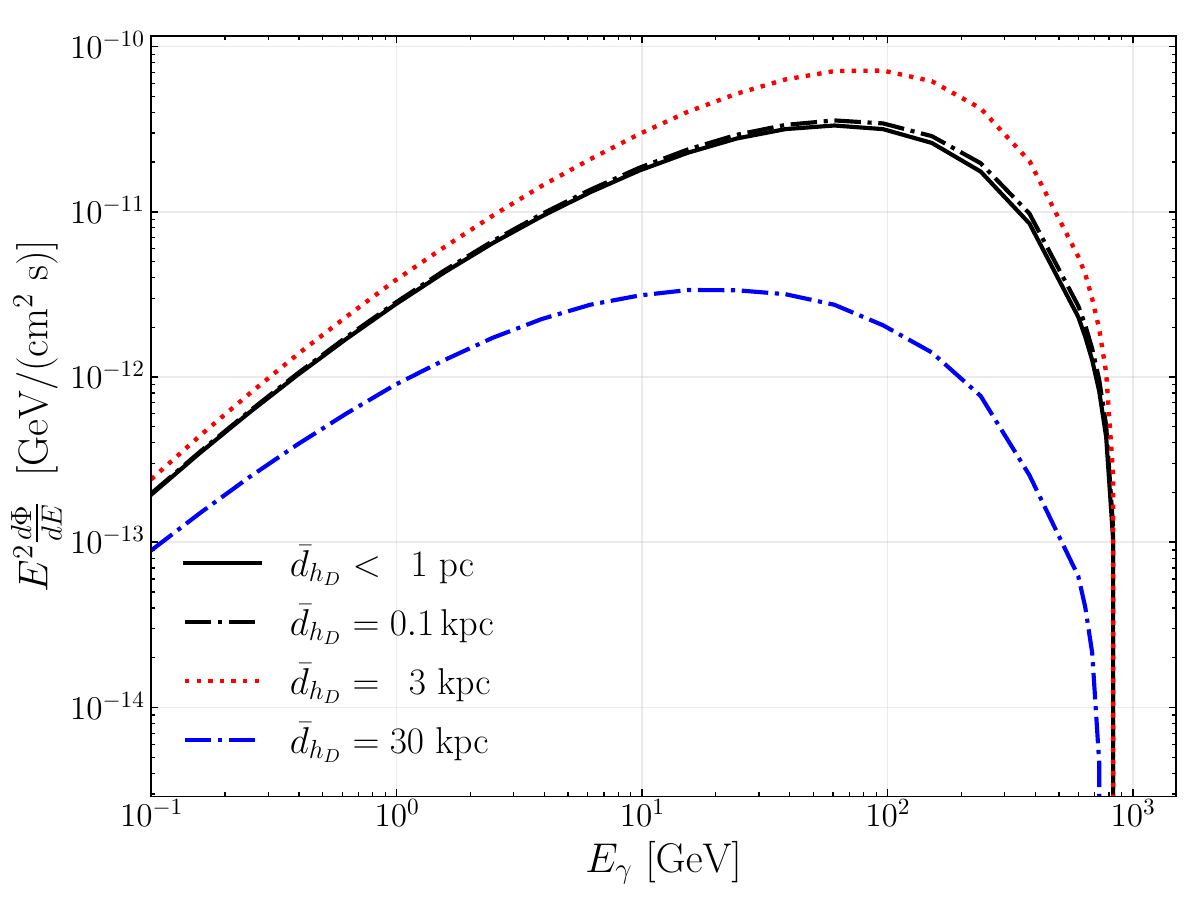}
\caption{The differential flux of photons produced in the DM-induced cascade annihilation process described in the text for the model under study is shown as a function of the photon energy. In the figure, we show the result for the standard scenario with the black solid line. In this case, the dark Higgs boson decays promptly after being produced in the $2\to 3$ annihilation process of $\chi$ DM. Instead, the black dash-dotted, red dotted, and blue dashed lines correspond to the long-lived mediator case with the relevant typical decay length of boosted $h_D$ equal to $0.1$, $3$, and $30$~kpc, respectively.
\label{fig:spectra}
}
\end{figure}

An example of such a photon spectrum is shown with the black solid line in \cref{fig:spectra}. We have obtained it for the fixed masses of the dark sector species $m_{\chi} = 1.5~\tev$, $m_{\eta} = 250~\gev$, $m_{h_D} = 5~\gev$, and $\Delta_{h_D A^\prime} = 0.02$. The coupling constants were fixed as follows: $\epsilon=10^{-6}$, $\alpha_D = 0.1$, and $\lambda_{h_D \eta}=0.5$. As expected, even though the assumed $\chi$ mass is above \tev, the resulting photon spectrum is much softer and peaked at $E_\gamma\sim 100\gev$.

In \cref{fig:spectra}, we further present several such photon spectra corresponding to very long-lived mediators with the decay length of order $\bar{d}_{h_D} \sim 0.1$, $3$ and $30~\textrm{kpc}$. We present the photon fluxes for the $|b|,|l|<12^\circ$ region around the GC. As discussed in \cref{app:nonlocal}, for $\bar{d}_{h_D}\ll d_{\textrm{RoI}}\simeq 2.3~\textrm{kpc}$ the impact of non-local effects on the observed spectrum is very small and the photon spectrum resembles the one obtained in the short lifetime regime. This is indicated by the black dash-dotted line in the figure. Instead, for larger values of the decay length, $\bar{d}_{h_D}\sim d_{\textrm{RoI}}$, we see a relative increase in the photon flux observed from this RoI due to anisotropy effects, as shown with the red dotted line. Finally, for very large decay lengths, $\bar{d}_{h_D}\gg d_{\textrm{RoI}}$, the flux becomes suppressed, which is illustrated with a blue dash-dotted line. The suppression is more pronounced for energetic photons as they originate from more boosted dark vectors. These can more easily escape the RoI before decaying. As a result, the observed photon spectrum is even more shifted towards lower energies.

%%%%%%%%%%%%%%%%%%%%%%%%%%%%%%%%%%%%%%%%%%%%%%%%%%%%%%%%%
\section{Non-local effects in $\gamma$-ray DM ID for $\chi$\label{app:nonlocal}}
%%%%%%%%%%%%%%%%%%%%%%%%%%%%%%%%%%%%%%%%%%%%%%%%%%%%%%%%%

As discussed in the main text, the phenomenology of the type of model of our interest differs from both traditional WIMPs and light DM species. Especially interesting such effects can be expected in indirect DM searches where secluded heavy DM can be detected thanks to non-negligible annihilation rates into lighter dark sector species. Importantly, the light BSM species can even be very long-lived and constrained by cosmology. This opens up new detection prospects for this kind of LLPs in the lifetime regime that remain highly complementary to intensity frontier searches.

In DM ID searches, this can lead to striking non-local effects~\cite{Rothstein:2009pm,Kim:2017qaw,Chu:2017vao,Gori:2018lem,Agashe:2020luo}. In this case, the interplay between different methods of performing DM ID observations could be challenging to interpret in standard WIMP scenarios, while it could indicate the existence of a dark sector comprising both heavy WIMP-like DM particles and much lighter and long-lived species. Additional non-local effects that can modify signals include enhanced DM-induced signal rates from extensive regions outside the Galactic Center (GC) and decreased signals from individual small regions, e.g., around dwarf spheroidal galaxies (dSphs). Below, we first discuss these effects for general WIMP DM and then present the possible impact on the parameter space of the model under study.

In the non-local case, the DM-induced $\gamma$-ray flux is partially driven outside the dense region around the GC by a very long-lived mediator $h_D$. As a result, both the energy spectrum and morphology of the signal might be affected. The respective photon flux reads
\begin{widetext}
    \begin{align}
        \label{eq:fluxnonlocal}
        \left(\frac{d\Phi}{dE_\gamma}\right)\Bigg|_{\textrm{non-local}} = \sum_{\textrm{bins}\,E_{h_D}} &\Bigg[\frac{1}{8\,\pi}\frac{\langle\sigma v\rangle_{E_{h_D}}}{m_{\chi}^2}\frac{1}{\bar{d}_{h_D}}\,\int_{\Delta\Omega}{d\Omega}\int_{\textrm{l.o.s.}}{ds}\int_{V_\chi}{d^3\vec{r}_{\chi}} \frac{\rho_{\chi}^2(|\vec{r}_{\chi}-\vec{r}_{\textrm{GC}}|)}{|\vec{r}_{h_D}-\vec{r}_{\chi}|^2}\,\\
        & \times\,\exp{\left(-\frac{|\vec{r}_{h_D}-\vec{r}_{\chi}|}{\bar{d}_{h_D}}\right)}\,\gamma_{h_D}\left(1-\beta_{h_D}\cos\theta\right)\,\frac{f(\theta)}{4\pi}\left(\frac{dN_\gamma}{dE_\gamma}\right)_{\chi}\Bigg|_{E_{h_D}}\Bigg]\nonumber
    \end{align}
\end{widetext}
where $\bar{d}_{h_D} = c\tau_{h_D}\gamma_{h_D}\beta_{h_D}$ is the decay length of $h_D$ in the Galactic frame and vectors $\vec{r_{\chi}}$, $\vec{r}_{h_D}$, and $\vec{r}_{\textrm{GC}}$ correspond, respectively, to the position of the $\chi$ annihilation, the $h_D$ decay, and the GC with respect to the detector on Earth. As can be seen, compared with the standard case, \cref{eq:flux}, in the non-local DM ID an additional integration appears over the position of the initial $\chi$ annihilation. This takes into account the fact that the long-lived mediator $h_D$ produced in $2\to 3$ processes at $\vec{r}_{\chi}$ can travel long-distances before decaying at position $\vec{r}_{h_D}$. In particular, the initial position $\vec{r}_{\chi}$ can lie outside the region of interest (RoI) in a given DM ID analysis. The mediator decay probability decreases exponentially with the growing distance $|\vec{r}_{h_D}-\vec{r}_{\chi}|$. Hence, typically only a limited region in the Galaxy around the RoI contributes to the observed DM-induced photon flux, although this depends on the value of $\bar{d}_{h_D}$.

In \cref{eq:fluxnonlocal}, we also employ anisotropy factors that depend on the angle $\theta$ defined as the angle between the $h_D$ boost direction and the detector,
\begin{equation}
\cos\theta = \frac{\vec{r}_{h_D}\cdot (\vec{r}_{\chi}-\vec{r}_{h_D})}{|\vec{r_{h_D}}||\vec{r}_{\chi}-\vec{r}_{h_D}|}.
\label{eq:costheta}
\end{equation}
In the case of a two-body decay of $h_D\to XX$, the function $f(\theta)$ reads
\begin{equation}
f(\theta) = \frac{(1+\tan^2\theta)^{3/2}}{\tan^2\theta}\,\frac{\left[(\beta/\tilde{\beta}_{X})+\cos\tilde{\theta}\right]\,\sin^2\tilde{\theta}}{(\beta/\tilde{\beta}_{X})\cos\tilde{\theta}+1},
\label{eq:ftheta}
\end{equation}
in which $\tilde{\theta}$ is the relevant angle in the $h_D$ rest frame and we obtain $\cos\tilde{\theta} = \cos\tilde{\theta}_+$ for $\theta\leq \pi/2$ and $\cos\tilde{\theta} =  \cos\tilde{\theta}_-$ otherwise, where
\begin{equation}
\cos\tilde{\theta}_{+,-} = \frac{-\gamma^2\tan^2\theta\,\frac{\beta}{\tilde{\beta}_{X}}\pm \sqrt{\gamma^2\tan^2\theta\,\left(1-\frac{\beta^2}{\tilde{\beta}_{X}^2}\right)+1}}{\gamma^2\tan^2\theta + 1}\ ,
\label{eq:costhposneg}
\end{equation}
and $\tilde{\beta}_{h_D} = \sqrt{1-(2m_{h_D}/m_{A^\prime})^2}$. In the simplest case (not relevant to our model), in which the traveling mediator directly decays into a pair of photons, $X=\gamma$, one would reproduce a known expression for \textsl{radiative beaming}, $f(\theta) = \gamma_{h_D}^2\,(\beta_{h_D}\cos{\tilde{\theta}}+1)^2$. The anisotropy factors appear due to the boost of the decaying $h_D$ in the Galactic frame. They affect the final observed photon flux at Earth since decaying mediators preferentially come from the direction of the GC in each given region in the Galaxy. However, in the non-relativistic and local limit, we obtain $f(\theta)\to 1$.

\begin{figure*}[tb]
\centering
\includegraphics[scale=0.4]{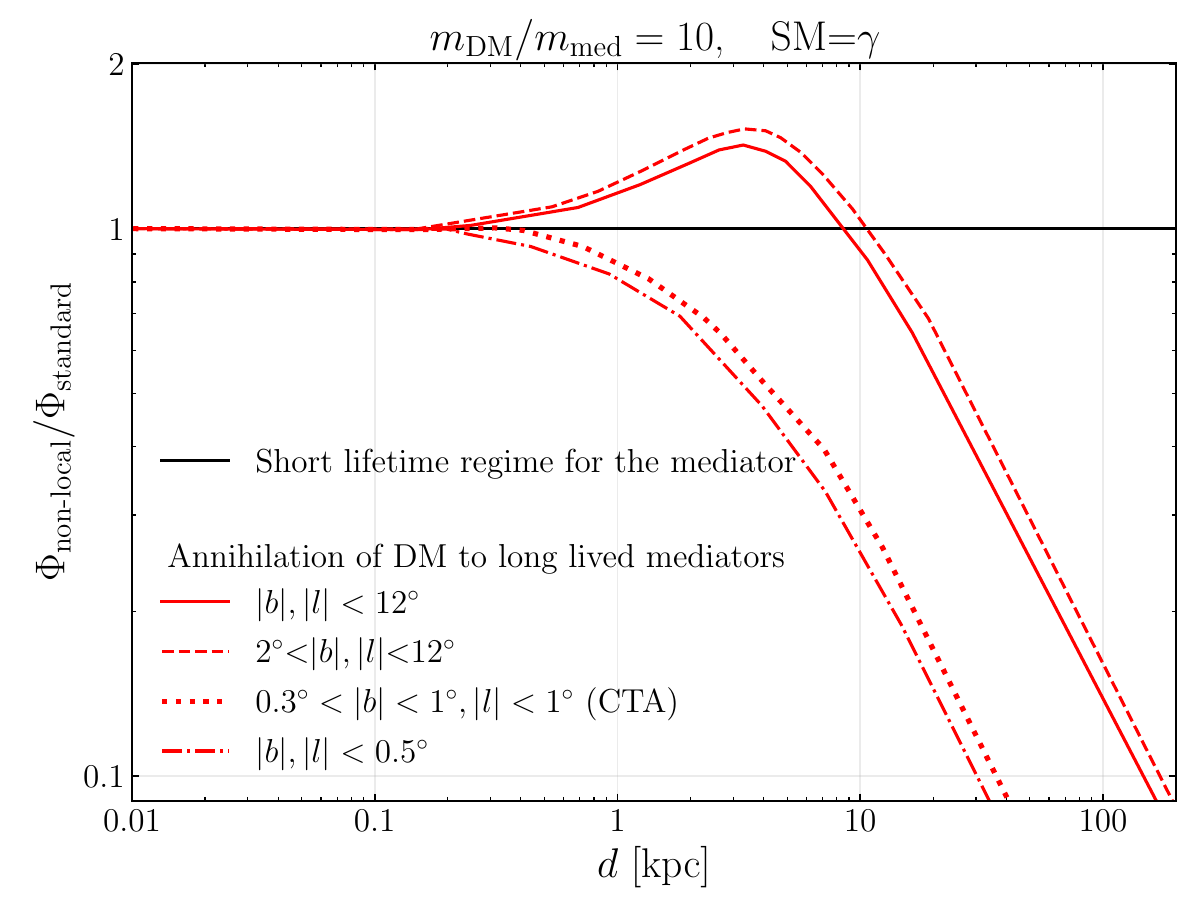}\hspace{0.3cm}\includegraphics[scale=0.41]{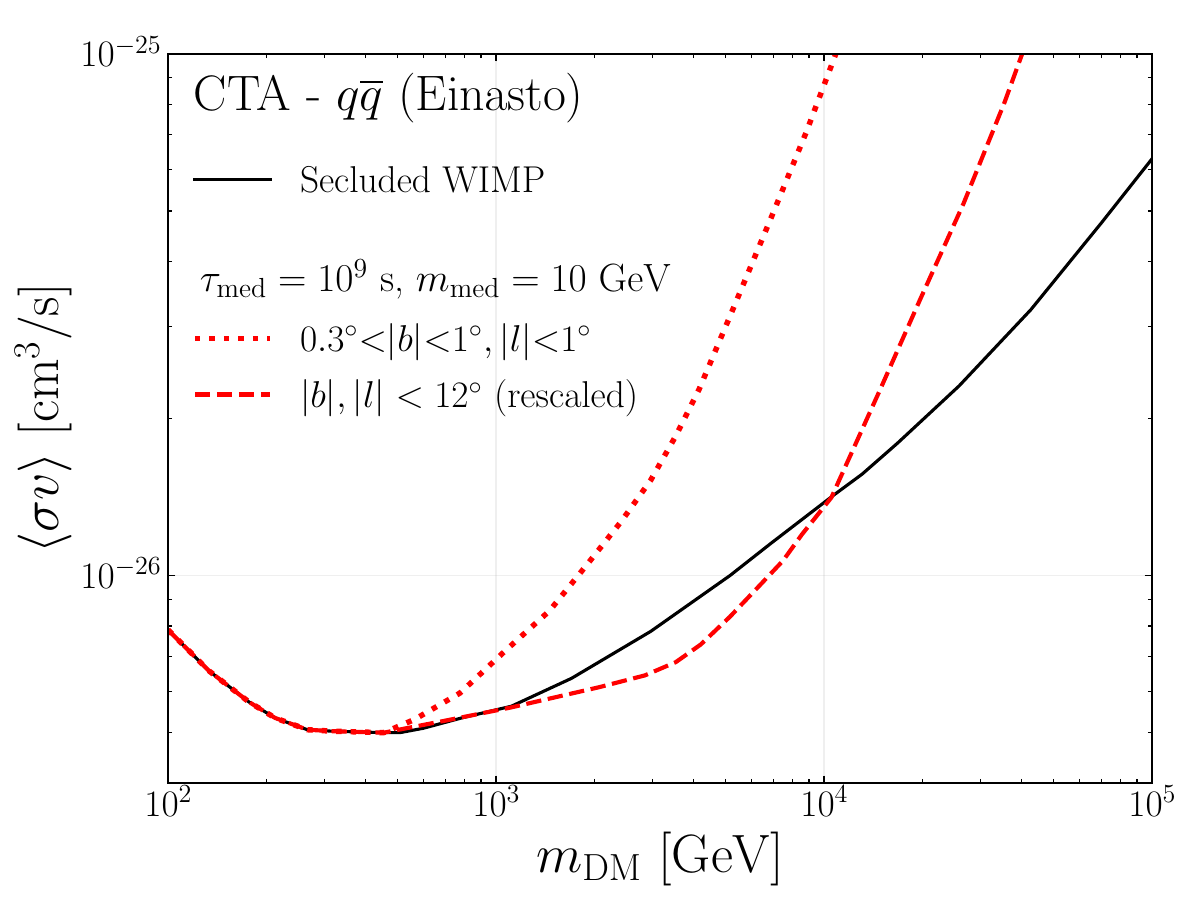}
\caption{Left: The ratio of the integrated photon fluxes obtained for increasing decay length $d$ of the mediator and in the standard regime of prompt decays shown with the horizontal black solid line. The figure has been prepared for the toy model with the DM and mediator masses equal to $m_{\textrm{DM}} = 100~\gev$ and $m_{\textrm{med}} = 10~\gev$, respectively, and assuming that the mediator decays directly into a pair of photons. We integrate the fluxes over the energy range between $0.1\gev$ and $100\gev$. The solid (dashed, dotted, dash-dotted) red line corresponds to the DM ID observation region around the GC defined by different longitude and latitude limits, as indicated in the figure. Right: The CTA sensitivity for the secluded WIMP DM scenario in the ($m_{\textrm{DM}}$, $\langle\sigma v\rangle$) plane is shown with the black solid line following Ref.~\cite{Siqueira:2021lqj}. For comparison, we also present with the red dotted line the expected such reach for the toy model with the long-lived mediator with $\tau_{\textrm{med}} = 10^9\second$ and $m_{\textrm{med}} = 10~\gev$ which decays into light quarks. The red dashed line corresponds to a re-scaled sensitivity for a larger RoI, as indicated in the figure (see text for details).
\label{fig:fluxes}
}
\end{figure*}

We illustrate the impact of non-local effects on the total integrated flux of $\gamma$-rays in a toy DM model with $m_{\textrm{DM}} = 100~\gev$ and a long-lived mediator mass $m_{\textrm{med}} = 10~\gev$ in the left panel of \cref{fig:fluxes}. Here, we assume for simplicity that the mediator decays directly into a pair of photons. We show with the red lines the ratio between the flux obtained in the long and short mediator lifetime regimes for several different RoIs. In particular, the solid red line corresponds to a large region around the GC characterized by $|b|,|l|<12^\circ$.\footnote{We use the Galactic coordinate system with the Galactic longitude $l$ and latitude $b$.}  While this is a larger region than for typical CTA analyses, we employ it to highlight the difference between the searches better focused on the closed vicinity of the GC and the possible extended Galactic center survey present in the non-local DM ID regime; see, e.g., Refs~\cite{CTAConsortium:2017dvg,CTA:2020qlo} for further discussion about CTA analyses. The large RoI employed in our study extends to roughly $d_{\textrm{RoI}} \sim R_0\,\sin{b} \simeq 2.3~\textrm{kpc}$ distance away from the GC, where $R_0 = 9~\textrm{kpc}$ is the distance between the Earth and the GC.

As can be seen in the figure, for $\bar{d}_{\textrm{med}}\lesssim d_{\textrm{RoI}}$ the impact of non-local effects on the observed spectrum is very small. The photon spectrum resembles the one obtained in the short lifetime regime denoted by the horizontal black solid line, $\Phi_{\textrm{nonlocal}}/\Phi_{\textrm{stand.}}\simeq 1$. In contrast, for very large decay lengths of $A^\prime$ the expected DM-induced photon flux coming from the RoI drops down much below the standard expectations. This is due to the efficient escape of mediators away from the GC before they decay. The decrease of the flux is roughly linear in growing $\bar{d}_{\textrm{med}}$, as can be seen from \cref{eq:fluxnonlocal} in the limit of $\bar{d}_{\textrm{med}}\gg |\vec{r}_{\textrm{med}}-\vec{r}_{\chi}|\sim d_{\textrm{RoI}}$.

The relative increase of the DM-induced photon flux for intermediate values of $\bar{d}_{\textrm{med}}\sim d_{\textrm{RoI}}$ can be understood as follows~\cite{Chu:2017vao}. In this case, most of the mediators produced close to the GC decay within the RoI. Only a small fraction of dark photons produced at the GC would generate photons traveling toward the Earth. In the standard case, the signal from other mediators will be lost. However, dark mediators can partially overcome this in the non-local regime by traveling away from the GC before decaying. At these remote positions, they can produce photons in the direction of the Earth, which would not be seen had they been produced close to the GC. As a result, the DM ID signal rates from distant positions within the RoI receive contributions from both local DM annihilations and annihilations occurring close to the GC. This increase is even more pronounced for a modified RoI around the GC in which we exclude the innermost region with the $2^\circ$ size aperture. We present this with the dashed red line in the figure. In this case, the photon flux in the non-local regime gains even more from the dark mediators traveling inside the RoI from the innermost region close to the GC.

In addition, we also show in the figure the expected photon fluxes for much smaller RoIs. Here, with a red dotted line, we present the results for a small RoI characteristic for the CTA Galactic center survey, in which we have additionally excluded part of the region very close to the GC, \ie, we assume $0.3^\circ < |b| < 1^\circ$ and $|l|<1^\circ$. Instead, with a red dash-dotted line, we present the flux for an even smaller region around the GC with $|b|,|l| < 0.5^\circ$. The size of this region encompasses a typical DM halo size for dwarf galaxies in the Fermi-LAT analyses~\cite{Fermi-LAT:2015att}. As can be seen, for both small RoIs, the relative growth of the flux for smaller $\bar{d}_{\textrm{med}}$ is hard to reconstruct, while for the decay length of order several kpcs, the flux is already suppressed. We note that if dSphs are modeled as point-like sources in the analysis with the $0.1^\circ\times 0.1^\circ$ bin size~\cite{Fermi-LAT:2016uux}, the impact of non-local effects is even stronger. Last but not least, we stress that in the non-local regime, the DM-induced photons escaping from small RoIs could also affect the analysis based on local background expectations around each of the dSphs. This could further ameliorate the relevant bounds.

As can be seen, for $\bar{d}_{\textrm{med}}\sim \textrm{a few kpc}$, the difference between the impact of non-local effects for DM ID focusing on extensive regions around the GC and for searches targeting small RoIs can reach up to a factor of a few in the predicted photon flux. One can then expect a weakening of the relevant DM constraints derived based on stacked dwarf analyses, while DM-induced signals could be stronger for searches using larger RoIs around the GC. This might open new possibilities in explaining persisting anomalies in DM searches; see, e.g., Ref.~\cite{Agashe:2020luo} for the relevant discussion about the Galactic Center Excess (GCE) and non-local DM ID effects. However, as mentioned above, such effects would also modify the morphology of the DM-induced signals, which could then no longer follow the original DM profile but appear less cuspy. In particular, as shown in Ref.~\cite{Agashe:2020luo}, when compared to the standard short-lived regime, DM solutions to the GCE employing non-local effects struggle to improve the global  $\chi^2$ fit for this anomaly. 

In the right panel of \cref{fig:fluxes}, we illustrate the impact of non-local effects on DM ID searches for the aforementioned toy model. To this end, we compare the expected sensitivity reach of the CTA in a secluded WIMP DM scenario presented in Ref.~\cite{Siqueira:2021lqj} with the relevant reach obtained for a very long-lived mediator with $\tau_{\textrm{med}} = 10^9\second$ and fixed $m_{\textrm{med}} = 10~\gev$. Here, we assume that the mediator decays into light quarks. In the figure, larger values of the mass of annihilating DM $m_{\textrm{DM}}$ imply larger values of the boost factor of the mediator and the corresponding decay length $\bar{d}_{\textrm{med}}\simeq (m_{\textrm{DM}}\,/\,1~\tev)\,(10~\gev/m_{\textrm{med}})\times 1~\textrm{kpc}$. This results in an effective suppression of the DM-induced signal from a small RoI around the GC for $m_{\textrm{DM}}\gtrsim \textrm{a few hundred}~\gev$, as indicated with the dotted red line in the figure. We also show there with the red dashed line the expected sensitivity for a larger RoI $|b|,|l|<12^\circ$, which, for illustrative purposes, has been re-scaled to match the sensitivity of the small RoI in the limit of low $m_{\textrm{DM}}$. As can be seen, for the extended RoI, the weakening of the future bounds corresponds to larger DM masses than for small RoI. In this case, we also observe a relative improvement of the bound in the intermediate region of $m_{\textrm{DM}}\sim \textrm{a few}~\tev$. This is due to the excess DM-induced photon flux for $\bar{d}_{\textrm{med}}\sim d_{\textrm{RoI}}$, as discussed above.

\begin{figure*}[tb]
    \centering
    \includegraphics[scale=0.38]{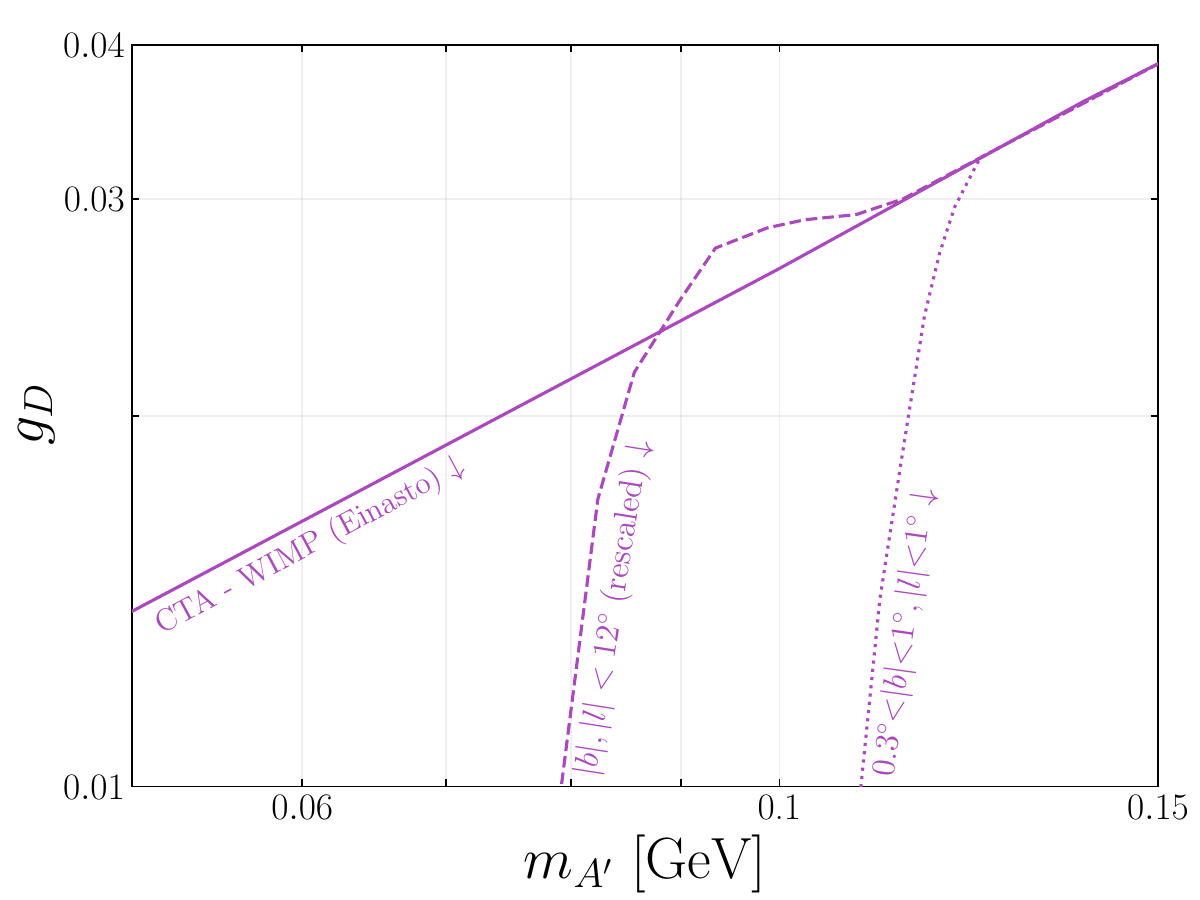}\hspace*{0.3cm}\includegraphics[scale=0.38]{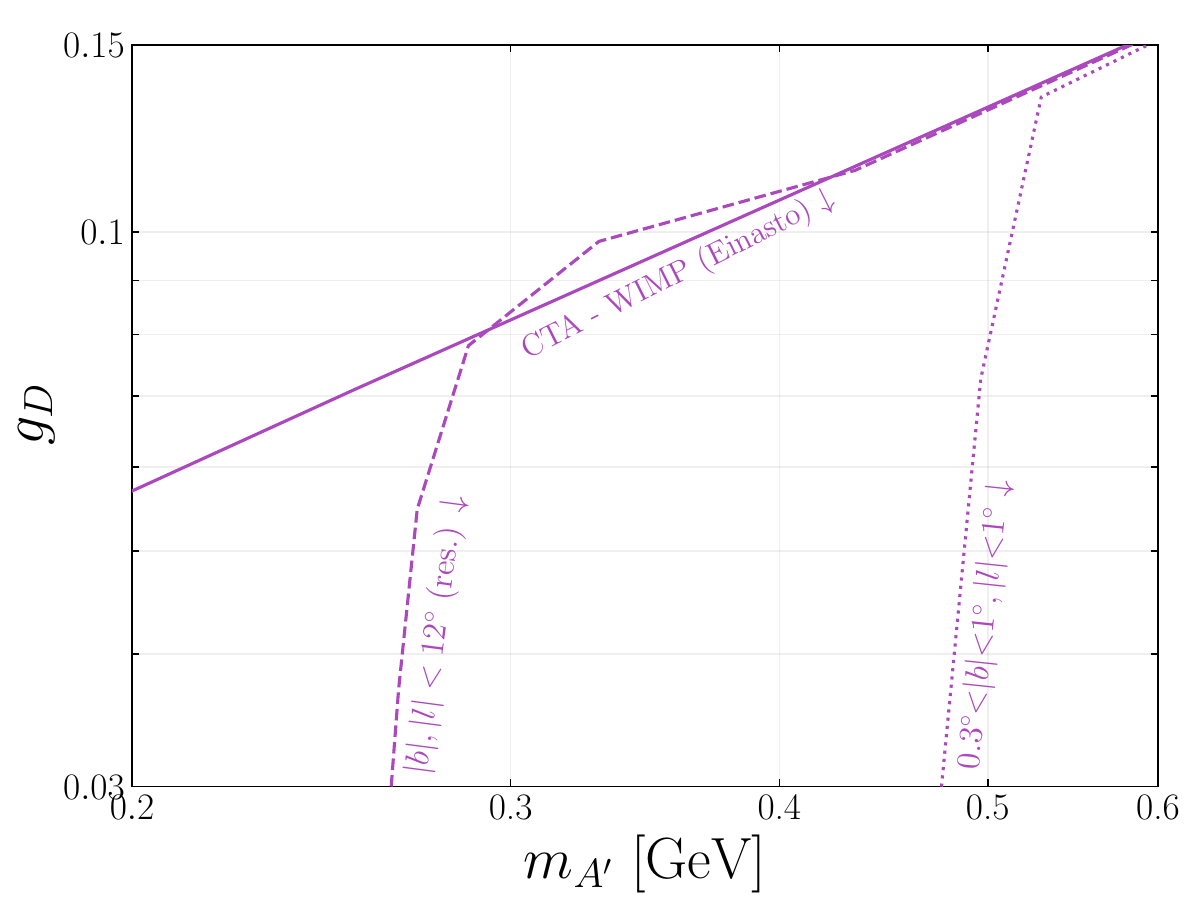}
    \caption{
    Similar to \cref{fig:results_mDM_gD_1} but here, zoomed-in plots are shown focusing on DM indirect detection bounds obtained based on $\gamma$-ray search toward the Galactic Center. In the plot, we fix $\lambda_{h_D \eta} =4\pi$ and assume $m_\chi=1.5\tev$ ($15\tev$) in the left (right) panel. The solid line corresponds to the projected CTA bound obtained without considering non-local effects. Dashed and dotted lines illustrate the impact of non-local effects on analyses for two different regions of interest around the GC, as indicated in the plots.
    \label{fig:results_nonlocal_ourmodel}
    }
\end{figure*}

In \cref{fig:results_nonlocal_ourmodel}, we show the results of a similar analysis of the BSM model described in the main text. In the left panel, we fix the heavy DM mass to $m_{\chi} = 1.5\tev$, while it is increased to $15\tev$ in the right panel. The solid purple lines in the plots correspond to projected CTA bounds obtained when neglecting non-local effects. We also show in the plots dashed and dotted lines obtained for two different RoI, as discussed above, for which the impact of increasing $h_D$ decay length is considered. As can be seen, for small values of $m_{A^\prime}\simeq m_{h_D}$ in the plot, which assumes $\Delta_{h_D A^\prime} = 0.01$, the projected CTA bounds become much weaker. This is expected based on the growing escape probability of $h_D$s outside the RoI. The same effect is also shown in \cref{fig:results_mDM_gD_1} in the main text. One also finds the aforementioned enhanced signal rate effect in the plots for selected values of the dark sector masses. 

Comparing left and right panels, we see that the non-local effects become important at even larger $m_{A^\prime}\simeq m_{h_D}$ for increasing $m_\chi$. This is because more boosted light dark Higgs bosons are then produced in the $2\to 3$ annihilation process, $\chi\chi\to\eta\eta h_D$. In addition, all the DM ID projected bounds are shifted to higher values of $g_D$ in this case. This is primarily driven by an increase in the typical energy carried away by $h_D$ after $\chi$ annihilation, which corresponds to the region with an improved CTA sensitivity. The negative impact of growing $g_D$ on the dark Higgs boson lifetime in \cref{eq:ctauhD} partially compensates the effect of increasing the boost factor of $h_D$ in estimating the non-local effects. As a result, the shift in $m_{A^\prime}\simeq m_{h_D}$ is lower than a naively expected order of magnitude difference driven by the change in $m_\chi$.

\bibliography{out.bib}
\end{document}